\documentclass[superscriptaddress,twocolumn,pra]{revtex4-1}

\usepackage{graphicx}
\usepackage{amsmath}
\usepackage{longtable}
\usepackage{amssymb}
\usepackage{latexsym}
\usepackage{natbib}
\usepackage{color}

\usepackage{times}
\usepackage{graphics}
\usepackage{epsfig}
\usepackage{wrapfig}
\usepackage[T1]{fontenc}
\usepackage{float}
\usepackage{enumerate}
\usepackage[compact]{titlesec}
\usepackage{setspace,ifthen}
\usepackage{amsthm} 
\usepackage{amssymb}	
\usepackage{times}
\usepackage[usenames,dvipsnames]{xcolor}
\usepackage{mathrsfs,amsmath}
\usepackage{mathrsfs}
\usepackage{mathrsfs}
\usepackage{rotating}
\usepackage{epstopdf}
\usepackage{appendix}

\usepackage{array}

\usepackage{rotate}
\usepackage{kbordermatrix} 

\renewcommand{\kbldelim}{(} 
\renewcommand{\kbrdelim}{)}

\newcommand{\abs}[1]{\left| #1 \right|} 
 
\let\baraccent=\= 
\renewcommand{\=}[1]{\stackrel{#1}{=}} 

\theoremstyle{definition}

\theoremstyle{remark}

\newcommand{\bla}{\color{black}}

\newcommand{\X}{\hat{X}}
\newcommand{\Y}{\hat{Y}}
\newcommand{\Z}{\hat{Z}}

\newcommand{\Pauli}[1]{\hat{\sigma}_{#1}} 
\newcommand{\LeviCevita}[1]{\epsilon_{#1}}

\newcommand{\RxPlus}{{R_x^{+}}}
\newcommand{\RyPlus}{{R_y^{+}}}
\newcommand{\RzPlus}{{R_z^{+}}}
\newcommand{\RxMin}{{R_x^{-}}}
\newcommand{\RyMin}{{R_y^{-}}}
\newcommand{\RzMin}{{R_z^{-}}}

\newcommand{\PauliGroup}{\mathcal{P}}
\newcommand{\Clifford}{\mathcal{\hat{C}}}

\newcommand{\Crun}[2][1]{C_{#2}}
\newcommand{\Kstring}[2][1]{K_{#2}}
\newcommand{\PTwirl}[2][1]{\mathbf{P}_{#2}}

\newcommand{\RBseqClean}{\mathcal{S}}
\newcommand{\RBseqNoisy}{\tilde{\mathcal{S}}}

\newcommand{\CliffVec}{\boldsymbol{\eta}}
\newcommand{\NoiseVec}{\boldsymbol{\delta}}
\newcommand{\CliffIndex}[2][1]{\eta_{#2}}

\newcommand{\RBTaylorTerm}[2]{\xi_{#2}^{(#1)}}
\newcommand{\HTRBTaylorTerm}[2]{\Q_{#2}^{(#1)}}
\newcommand{\ProbFail}{\delta}

\newcommand{\UniError}[2]{\Delta^{(#2)}_{#1}}
\newcommand{\UniErrorVec}{\vec{\boldsymbol{\Delta}}}
\newcommand{\RBTaylorTermUni}[2]{\tilde{\xi}_{#2}^{(#1)}}
\newcommand{\HTRBTaylorTermUni}[2]{\tilde{\Q}_{#2}^{(#1)}}

\newcommand{\sinc}{\text{sinc}}
\newcommand{\FT}{\mathscr{F}}

\newcommand{\Identity}{\mathbb{I}}
\newcommand{\Order}[2][1]{\mathcal{O}\left(#2\right)}
\newcommand{\Trace}[2][1]{\text{Tr}\left(#2\right)}

\newcommand{\FidNoiseAv}{\left\langle\mathcal{F}\right\rangle}

\newcommand{\Fid}[1]{F_{#1}}

\newcommand{\E}[2][1]{\mathbb{E}[#2]}
\newcommand{\Var}[2][1]{\mathbb{V}[#2]}

\newcommand{\Q}{Q}

\newcommand{\norm}[1]{\| #1 \|} 

\newcommand{\eqn}[1]{Eq.~\ref{#1}}

\newcommand{\Mmax}{M_\textrm{max}}

\newcommand{\rhat}{\hat{\boldsymbol{r}}}
\newcommand{\xhat}{\hat{\boldsymbol{x}}}
\newcommand{\yhat}{\hat{\boldsymbol{y}}}
\newcommand{\zhat}{\hat{\boldsymbol{z}}}

\newcommand{\RvecA}{\vec{\boldsymbol{R}}}
\newcommand{\Vvec}{\vec{\boldsymbol{V}}}

\newcommand{\RnormSQ}{\norm{\RvecA}^2}

\begin{document}



\title{The effect of noise correlations on randomized benchmarking}

\author{Harrison Ball}
\affiliation{ARC Centre of Excellence for Engineered Quantum Systems, School of Physics, The University of Sydney, NSW 2006 Australia}
\affiliation{Australian National Measurement Institute, West Lindfield, NSW 2070}
\author{Thomas M. Stace}
\affiliation{ARC Centre of Excellence for Engineered Quantum Systems, School of Physics and Mathematics, The University of Queensland, QLD 4072 Australia}
\author{Steven T. Flammia}
\affiliation{ARC Centre of Excellence for Engineered Quantum Systems, School of Physics, The University of Sydney, NSW 2006 Australia}\author{Michael J. Biercuk}
\email[Correspondence to: ]{michael.biercuk@sydney.edu.au}
\affiliation{ARC Centre of Excellence for Engineered Quantum Systems, School of Physics, The University of Sydney, NSW 2006 Australia}\affiliation{Australian National Measurement Institute, West Lindfield, NSW 2070}

\date{\today}



\begin{abstract}
 Among the most popular and well studied quantum characterization, verification and validation techniques is randomized benchmarking (RB), an important statistical tool used to characterize the performance of physical logic operations useful in quantum information processing.  In this work we provide a detailed mathematical treatment of the effect of temporal noise correlations on the outcomes of RB protocols.  We provide a fully analytic framework capturing the accumulation of error in RB expressed in terms of a three-dimensional random walk in ''Pauli space.''  Using this framework we derive the probability density function describing RB outcomes (averaged over noise) for both Markovian and correlated errors, which we show is generally described by a gamma distribution with shape and scale parameters depending on the correlation structure.  Long temporal correlations impart large nonvanishing variance and skew in the distribution towards high-fidelity outcomes -- consistent with existing experimental data -- highlighting potential finite-sampling pitfalls and the divergence of the mean RB outcome from worst-case errors in the presence of noise correlations.  We use the Filter-transfer function formalism to reveal the underlying reason for these differences in terms of effective coherent averaging of correlated errors in certain random sequences.  We conclude by commenting on the impact of these calculations on the utility of single-metric approaches to quantum characterization, verification, and validation.

\end{abstract}

\pacs{}

\maketitle


\section{Introduction}

Quantum verification and validation protocols are a vital tool for characterizing quantum devices. These take many forms~\cite{WhiteStateTomography,WhiteProcessTomography, Emerson2005, Cory2007, Pablo2008, CoryCorrelations, Walther2013, Magnesan2013}, but one of the most popular due to its efficiency is randomized benchmarking (RB).  In this approach, developed originally by Knill et al.~\cite{Knill2008}, and expanded theoretically by various authors~\cite{Emerson2011, MagesanInterleaved, Flammia2014, Epstein2014}, the average error probability of a quantum gate (\emph{e.g.}\ a bit flip) is estimated by implementing a randomly sampled gate sequence from the set of Clifford operations, and measuring the difference between the ideal transformation and the actual result.  Averaging over many randomized sequences yields information about the underlying gate fidelity.  

RB has become so important in the experimental community~\cite{Knill2008, BiercukQIC, Porto2010, GaeblerPRL2012, BrownPRA2011, ChowPRL2012, MartinisOptimized, Lucas2014, Saffman2015, Laflamme2015, Morello2015} that despite the experimental complexity it is now common to simply quote a measured gate error, $p_{\text{RB}}$, resulting from a RB measurement for a particular experimental system, and relate this number to tight bounds such as fault-tolerance error thresholds in quantum error correction~\cite{NC}.  Of late, reported values of $p_{\text{RB}}$ have compared favorably with these thresholds, and have been used to justify the scalability of particular experimental platforms.  Underlying this entire approach is the subtle but central assumption that errors are \emph{uncorrelated}~\cite{Knill2008}: an assumption which is generally violated by a wide variety of realistic error processes with long temporal correlations~\cite{BiercukQIC, Bylander2011, Clarke2013}.  Violation of this assumption has been noted previously as a cause of  increased variance of outcomes over randomizations, but until now there has not been a quantitative means to understand the impact of such temporal correlations, a detailed physical mechanism to explain why such distortion of RB results can appear, or a clear understanding of the impact of this observation on the applicability of RB.

In this manuscript we examine the impact of relaxing the Markovian-error assumption by studying its effect on the distribution of measurement outcomes over randomizations. We find that while all randomizations (meeting certain criteria) are valid within the RB framework, they exhibit starkly different susceptibility to error when those errors exhibit temporal correlations over multiple individual Clifford operations.  We provide a detailed mathematical treatment of error accumulation in RB, using a general model treated in the specific case of dephasing errors.   We demonstrate how the reduction of fidelity for a particular sequence may be given a geometric interpretation, taking the form of a random walk in a 3D Cartesian coordinate system.  The steps of this walk correspond to the appearance of Pauli $\X,\Y,\Z$ errors derived from interleaved dephasing operators in the sequence of Clifford operations, and the overall statistics are determined by the underlying correlations in the noise process.   Our treatment includes both extremal cases of uncorrelated (Markovian) and systematic (DC) errors, as well as generic correlations interpolating between these limiting cases and captured through a noise power spectral density.  Our results provide simple analytic forms for the probability density functions of fidelity outcomes over RB randomizations as a gamma distribution, building connections to engineering literature on failure analysis.  We describe the impact of these observations on the interpretation of RB measurement outcomes in various noise environments and highlight the disconnect between measured RB error and metrics relevant to fault-tolerance such as worst-case errors.

\vspace{\baselineskip}
\section{Randomized Benchmarking Procedure}\label{Sec:StandardRB}


A common experimental implementation of single-qubit RB involves measuring the fidelity of net operations composed from random sequences of Clifford operations.  For a sequence of length $J$, the net operation is written as the product 
\begin{equation}
\RBseqClean_{\CliffVec} \equiv \prod_{j=1}^{J}\Clifford_{\CliffIndex{j}}
\end{equation} of Clifford operators $\Clifford_{\CliffIndex{j}}$ (see Table \ref{Table:CliffordGroup}) indexed by the sequence $\CliffVec = (\CliffIndex{1},\CliffIndex{2},...,\CliffIndex{J})$, where the $\eta_{j}$ are random variables uniformly sampled from the set $\{1,2,...,24\}$ labelling the elements of the Clifford group. Typically one makes the constraint on $\CliffVec$ that, absent any error processes, the ideal RB sequence performs the identity operation $\RBseqClean_{\CliffVec} \equiv \Identity$. Of the $24^J$ total sequence combinations for a given $J$, only a small subset of $k\ll24^{J}$ random Clifford sequences is implemented in practice. Each sequence is repeated (with appropriate qubit reinitialization) $n$ times to build an ensemble of fidelity measurements sampling and the underlying error process. These measurement outcomes are then averaged together to obtain useful information from projective qubit measurements, with the mean fidelity retained for each random sequence.

We represent the space of measurement outcomes by the ($k\times n$) matrix
\begin{align}
\label{Eq:FiniteRBSampleSpaceArray}
\boldsymbol{F}^{(J)}
\equiv
\begingroup
  \renewcommand*{\arraystretch}{1.15}%
  \kbordermatrix{
           & \NoiseVec_{1}           & \NoiseVec_{2}           & \hdots           & \NoiseVec_{n}     \cr
    \CliffVec_{1}   &     \Fid{1,1}              & \Fid{1,2}        &\hdots       & \Fid{1,n} \cr
    \CliffVec_{2}    & \Fid{2,1}             & \Fid{2,2}       &\hdots      & \Fid{2,n} \cr
    \vdots & \vdots        & \vdots        & \ddots       &  \vdots  \cr
    \CliffVec_{k}    & \Fid{k,1}             & \Fid{k,2}       & \hdots       & \Fid{k,n} \cr
  }%
\endgroup
\end{align}
\noindent 
where element $\Fid{i,j}\in[0,1]$ is the measured fidelity when implementing the $i$th Clifford sequence \mbox{$\CliffVec_{i}= (\CliffIndex{i,1},\CliffIndex{i,2},...,\CliffIndex{i,J})$} in the presence of the $j$th realization of the error process, denoted by $\NoiseVec_{j}$. For clarity, we make the distinction between the \emph{measured} fidelity $\Fid{i,j}$ (\emph{e.g.} obtained in experiment via projective qubit measurements) and the \emph{calculated} fidelity (see Eq. \ref{Eq:TraceFidelityDefinition}) serving as a proxy for $\Fid{i,j}$ in our analytic framework. The process of repeating each $\CliffVec_i$ and averaging the fidelity outcomes  \emph{over the finite sample of noise realizations} therefore corresponds to a measurement of the ``noise-averaged'' fidelity for the $i$th sequence, represented by the quantity 
\begin{align}\label{Eq:SimulatedNAF}
\overline{\boldsymbol{F}}^{(J)}_{i,\langle\cdot\rangle}&\equiv \frac{1}{n}\sum_{j=1}^{n}\Fid{i,j},
\hspace{0.5cm}
i\in\{1,...,k\}
\end{align}
where the angle brackets $\langle\cdot\rangle$ in the second subscript denote averaging over the columns of $\boldsymbol{F}^{(J)}$.  Similarly the mean fidelity averaged over both Clifford sequences and errors is denoted 
\begin{align}\label{Eq:FidelityTotalSampleMean}
\hat{\mu}^{(J)}&\equiv\overline{\boldsymbol{F}}^{(J)}_{\langle\cdot\rangle,\langle\cdot\rangle} 
=\frac{1}{kn}\sum_{i=1}^{k}\sum_{j=1}^{n}\Fid{i,j}
\hspace{0.5cm}
\end{align}
where angle brackets in both subscripts indicate averaging over both rows and columns (that is, all elements) of $\boldsymbol{F}^{(J)}$. The quantity $\hat{\mu}^{(J)}$ is therefore an \emph{estimator} of the true mean fidelity $\mu^{(J)}\equiv\langle\mathcal{F}\rangle_{\CliffVec,\NoiseVec}$, formally obtained as the expectation over all possible fidelity outcomes $\mathcal{F}$ defined on the support of the random variables $\CliffVec$ and $\NoiseVec$. 

In the standard RB procedure, measurements of $\hat{\mu}^{(J)}$ for increasing $J$ are fitted to an exponential decay from which the mean gate error $p_{\text{RB}}$ is extracted as the decay constant. Implicit in this procedure, however, is the assumption that, for any random set of Clifford sequences, the resulting distribution of fidelity outcomes represents the underlying error process \emph{fairly}, and the total mean  $\hat{\mu}^{(J)}$ is reasonably representative of any individual sequence. That is, the distribution of values $\overline{\boldsymbol{F}}^{(J)}_{i,\langle\cdot\rangle}$ is symmetric about $\hat{\mu}^{(J)}$ with small relative variance for any random set $\{\CliffVec_{1},...,\CliffVec_{k}\}$. 

In this paper we show $\hat{\mu}^{(J)}$ to be an unbiased and effective estimator only when the error process is truly Markovian. That is, it possesses no temporal correlations.  We provide a detailed analytic calculation of the impact of realistic, correlated noise on the distribution of outcomes over different randomizations. To examine this we introduce a physical model compatible with the experimental procedure, and which permits accounting for the presence of noise correlations.

\vspace{\baselineskip}
\section{Physical model}\label{Sec:PhysicalModel}

\indent In this section we develop the mathematical framework used to model and investigate the impact of noise correlations on RB.  Our model assumes a Unitary error process with temporal correlations, reflecting typical experimental conditions where qubit rotations, \emph{e.g.} from a fluctuating classical field, dominate. This process is generically described by a power spectral density (PSD) capturing the various correlation timescales.  Our error model is fully general and considers \emph{universal} (multi-axis) errors.  For simplicity of technical presentation we treat single-axis dephasing in the main text and provide the full presentation of universal noise in Appendix \ref{Appendix:UniversalErrors}.  

The main result of this section is the geometric interpretation of error accumulation in RB sequences in terms of a random walk in 3 dimensions (a step taken for each Clifford), with step lengths and directionality inheriting the correlation structure of the error process. This result is independent of any choice of representation of the \emph{Clifford group}, though our numeric simulations later used to verify it do rely on a particular representation (see Table \ref{Table:CliffordGroup}).

\subsection{Noise-averaged fidelity $\FidNoiseAv$}\label{Sec:NoiseAveragedFidelity}
 
We begin by introducing our metric to quantify the reduction in fidelity for a given $\CliffVec$ and $\NoiseVec$.  We employ the trace fidelity 
\begin{align}
\label{Eq:TraceFidelityDefinition}
\mathcal{F}\left(\CliffVec,\NoiseVec\right)\equiv
\left|
\frac{1}{2}
\text{Tr}
\left(\RBseqClean_{\CliffVec}^\dagger\RBseqNoisy_{\CliffVec,\NoiseVec}
\right)
\right|^2
=
\frac{1}{4}
\left|
\text{Tr}\left(
\RBseqNoisy_{\CliffVec,\NoiseVec}
\right)
\right|^2
\end{align}
capturing the overlap between ideal, $\RBseqClean_{\CliffVec}$, and noise-affected sequences, $\RBseqNoisy_{\CliffVec,\NoiseVec}$, via the Hilbert-Schmidt inner product. For inputs $\CliffVec_i$ and $\NoiseVec_j$ this provides us with a computational metric corresponding to measurements $\Fid{i,j}$ obtained in experiment. Consequently our proxy for the measured noise-averaged fidelity $\overline{\boldsymbol{F}}^{(J)}_{i,\langle\cdot\rangle}$ takes the computational form 
\begin{align}\label{Eq:NoiseAveragedFidelityDefinition}
\left\langle
\mathcal{F}
\right\rangle_{\NoiseVec,n}
= \frac{1}{4}
\left\langle
\left|
\text{Tr}
\left(\RBseqNoisy_{\CliffVec,\NoiseVec}
\right)
\right|^2\right\rangle_{\NoiseVec,n}
\end{align}
where  $\langle\cdot\rangle_{\NoiseVec,n}$ denotes an ensemble average over $n$ realizations of $\NoiseVec$, and explicit reference to $\CliffVec$ and $\NoiseVec$ in the argument of $\mathcal{F}$ has been dropped. Henceforth we also drop the subscripts and denote the calculated noise-averaged fidelity simply by $\FidNoiseAv$. In the following sections we proceed with our main task: deriving the probability density function (PDF) of $\FidNoiseAv$, and its dependence on the correlation structure of $\NoiseVec$.

\subsection{Error model}\label{Sec:ErrorModel}

To reiterate, a Clifford sequence $\CliffVec$ implements the operation $\RBseqClean_{\CliffVec} \equiv \prod_{i=1}^{J}\Clifford_{\CliffIndex{i}}$, where $\Clifford_{\CliffIndex{i}}$ is the $\CliffIndex{i}^\text{th}$ Clifford generator. The random variables $\eta_{i}$ are uniformly sampled from the set $\{1,2,...,24\}$, and are assumed to be independent and identically distributed (i.i.d.), subject to the technical constraint $\RBseqClean_{\CliffVec} \equiv \Identity$. Unitary errors are implemented by interleaving $\RBseqClean_{\CliffVec}$ with a sequence of stochastic qubit rotations, yielding the noise-affected operation
\begin{align}
\label{Eq:RBNoisySequenceFullExpression}
\RBseqNoisy_{\CliffVec,\NoiseVec}&\equiv  U_1\bla\Clifford_{\CliffIndex{1}}  
 U_2\bla\Clifford_{\CliffIndex{2}}  
\bla...
U_J\bla\Clifford_{\CliffIndex{J}}.
\end{align}
This approach holds for arbitrary unitaries, $U_j$, enacting rotations in any Cartesian direction; a full treatment for universal noise models appears in Appendix~\ref{Appendix:UniversalErrors}. However, in the following we restrict ourselves to the subset of unitaries enacting a sequence of \emph{dephasing} rotations, $U_j\equiv \exp(-i\delta_j\Z)$,  parameterized by the time-series vector \mbox{$\NoiseVec = (\delta_{1},\delta_{2},...,\delta_{J})$}. We assume the error process is wide-sense stationary (\emph{i.e.} the for errors $\delta_i$ and $\delta_j$ the two-point correlation function $\langle\delta_{i}\delta_{j}\rangle=\langle\delta_{i}\delta_{i-k}\rangle$ depends only on the \emph{time difference} $k = i-j$), and has zero mean. 

Temporal noise correlations are established by introducing correlations between the elements of $\NoiseVec$.  In this work we treat three distinct cases
\begin{enumerate}
\item {\em Markovian process}:  Elements $\delta_j \sim\mathcal{N}(0,\sigma^2)$ are i.i.d. Gaussian-distributed errors with zero mean and variance variance $\sigma^2$, and completely uncorrelated between distinct Clifford gates in any sequence $\RBseqClean_{\CliffVec}$ (correlation length $1$).  
\item {\em DC process}:  Elements $\delta_j\equiv\delta\sim\mathcal{N}(0,\sigma^2)$ are identical over a given sequence $\RBseqClean_{\CliffVec}$ (maximally correlated, correlation length $J$), but are i.i.d. (uncorrelated) Gaussian-distributed errors with zero mean and variance variance $\sigma^2$ over different instances.
\item {\em Generically-correlated process}: Correlations between elements of $\NoiseVec$ separated by a time interval of ``$k$ gates'' in $\RBseqClean_{\CliffVec}$ are generically specified by an autocorrelation function $C_{\NoiseVec}(k) \equiv \left\langle \delta_j \delta_{j+k}\right\rangle$, in terms of which $\NoiseVec$ may be described by a power spectral density (PSD) $S(\omega)$.   
\end{enumerate}
As we will show, the noise interaction ``steers'' the operator product away from the identity gate performed by the ideal sequence $\RBseqClean_{\CliffVec}$ and reduces the operational fidelity in some way characteristic of the correlation structure of $\NoiseVec$.

\subsection{Analytic Expression for Sequence Fidelity}\label{Sec:WeakNoiseApproximation}

\subsubsection{Series expansion for Sequence Fidelity}

We begin by obtaining an approximation for $\FidNoiseAv$. As RB is intended to estimate errors too small to resolve in a single gate implementation, 
we assume the noise strength per gate, $\sigma$, is small.  Over long gate sequences the \emph{accumulation} of errors may be quantified by $J \sigma^2$, which can be large for sufficiently large $J$.  Assuming $J\sigma^2\lesssim1$, however, we can make analytical progress by expressing the error unitaries  in terms of a power series truncated at $\Order{\sigma^4}$. 
The noise-affected sequence Eq. \ref{Eq:RBNoisySequenceFullExpression} is therefore approximated by 
\begin{align}
\label{Eq:RBNoisySequenceTaylorExpandExpression1}
\hspace{-0.15cm}\RBseqNoisy_{\CliffVec,\NoiseVec}&\approx 
\prod_{j=1}^J\left(
\Identity 
+i\delta_{j}\Z
-\frac{\delta_{j}^2}{2}\Z^2 
-\frac{i\delta_{j}^3}{6}\Z^3
+\frac{\delta_{j}^4}{24}\Z^4
\right) \bla          \Clifford_{\CliffIndex{j}}
\end{align}
The full expansion of Eq. \ref{Eq:RBNoisySequenceTaylorExpandExpression1} generates products of order up to $\Order{
\prod_{j=1}^J
\delta_{j}^{4}
} = \Order{\sigma^{4J}}$. 
Let $\RBTaylorTerm{n}{k_1,k_2,...,k_m}$ denote the 
$\Order{
\prod_{\rho=1}^m
\delta_{j_\rho}^{k_\rho}
} = \Order{\sigma^{n}}$ 
product due to cross-multiplying terms like $(\delta_{j_1}\Z)^{k_1}(\delta_{j_2}\Z)^{k_2}... (\delta_{j_m}\Z)^{k_m}$, where $\sum_{\rho=1}^{m}k_\rho = n$. Retaining terms only up to fourth order ($n=4$), consistent with the order of our original Taylor approximation, we thereby obtain 
\begin{align}
\label{Eq:RBNoisySequenceTaylorExpandExpression2}
\RBseqNoisy_{\CliffVec,\NoiseVec}&\approx 
\RBTaylorTerm{0}{} 
+\RBTaylorTerm{1}{1} 
+\RBTaylorTerm{2}{1,1} 
+\RBTaylorTerm{2}{2} 
+\RBTaylorTerm{3}{1,1,1} 
+\RBTaylorTerm{3}{2,1} 
+\RBTaylorTerm{3}{3} \nonumber\\
&+\RBTaylorTerm{4}{1,1,1,1} 
+\RBTaylorTerm{4}{1,1,2} 
+\RBTaylorTerm{4}{2,2} 
+\RBTaylorTerm{4}{4}+\Order{\sigma^{6}}.
\end{align}
The first term in this expansion is identical to the ideal Clifford sequence, $\RBTaylorTerm{0}{} \equiv\RBseqClean_{\CliffVec}$. Higher-order terms capture successive error contributions and are composed of blocks, or subsequences, within $\RBseqClean_{\CliffVec}$ interrupted by one or more $\Z$ errors, indicated by the subscripts. To evaluate Eq. \ref{Eq:TraceFidelityDefinition} we must obtain expressions for the quantities $\HTRBTaylorTerm{n}{k_1,k_2,...,k_m}\equiv\frac{1}{2}\Trace{\RBTaylorTerm{n}{k_1,k_2,...,k_m}}$. Detailed derivations for these terms are given in Appendix~\ref{Appendix:ApproximatingFidNoiseAv}.

\subsubsection{Mapping from Clifford Sequence to ``Pauli space''}
We compute the terms in the power series by relating a given Clifford sequence to a random walk in ``Pauli space''.  For illustrative purposes,  we consider in detail only the quadratic terms:
\begin{align}\label{Eq:DominantXiTerm}
\RBTaylorTerm{2}{1,1}  &=-\sum_{j<k}\delta_j\delta_k\Crun{1,j-1}\Z\Crun{j,k-1}\Z\Crun{k,J}
\end{align}
where the ordered summand runs over $1\le j <k \le J$, and we have defined the Clifford \emph{subsequence} operators
\begin{align}
\Crun{jk}\equiv\Clifford_{\CliffIndex{j}}...\Clifford_{\CliffIndex{k}},
\hspace{0.75cm}
1\le j\le k\le J.
\end{align} 
We now define the cumulative operators $\Kstring{m}$ giving the product of the first $m$ Clifford operations
\begin{align}
\Kstring{m}\equiv\Crun{1,m} =
\Clifford_{\CliffIndex{1}}...\Clifford_{\CliffIndex{m}},
\end{align}
where $\Kstring{0}\equiv\Kstring{J}\equiv\Identity$, and \mbox{$\Kstring{m}^\dagger=\Kstring{m}^{-1}$} is also a  Clifford operator (since it is a product of Clifford operators). Any subsequence $\Crun{jk}$ therefore ``factorizes'' as
\begin{align}
\Crun{jk} =
\Clifford_{\CliffIndex{j-1}}^\dagger...\Clifford_{\CliffIndex{1}}^\dagger
\Clifford_{\CliffIndex{1}}...\Clifford_{\CliffIndex{j-1}}\Clifford_{\CliffIndex{j}}...\Clifford_{\CliffIndex{k}}
=
\Kstring{j-1}^\dagger
\Kstring{k} \bla
\end{align}
allowing us to rewrite 
\begin{align}\label{Eq:DominantXiOperatorsInTermsOfPtwirls}
\Crun{1,j-1}
\Z
\Crun{j,k-1}
\Z
\Crun{k,J}
=
\PTwirl{j}
\PTwirl{k}
\end{align}
where the operators on the right hand side are defined by 
\begin{align}
\label{Eq:DefinitionPTwirl}
\PTwirl{m}\equiv\Kstring{m-1}\Z\Kstring{m-1}^\dagger\in\{\pm\X,\pm\Y,\pm\Z\}
\end{align}
for $0\le m \le J$, and are always signed Pauli operators (since the Clifford group is the normalizer of the Pauli group). We may therefore express the operators $\PTwirl{m}$ in the basis of Pauli operators as
\begin{align}
\PTwirl{m} = x_m\X+y_m\Y+z_m\Z
\end{align}
where $x_m,y_m,z_m\in\{0,\pm1\}$ subject to the constraint $\abs{x_m}^2+\abs{y_m}^2+\abs{z_m}^2=1\nonumber$ (only one nonzero coefficient). The unit vector defined by 
\begin{align}
\rhat_{m} \equiv (x_m,\hspace{0.05cm}y_m,\hspace{0.05cm}z_m),\hspace{1cm}\|\rhat_m\| = 1
\end{align}
therefore points uniformly at random along one of the principle Cartesian axes $\{\pm\xhat,\pm\yhat,\pm\zhat\}$, and maps the ``direction'' of the operator $\PTwirl{m}$ in Pauli space.  Thus we have constructed a map from a given random Clifford sequence of length $J$, to a set of $J$ unit vectors each oriented at random along the cartesian axes.

With these insights the error contributions may be recast into more convenient expressions by moving to vector notation. In particular, taking the trace over Eq. \ref{Eq:DominantXiOperatorsInTermsOfPtwirls} and using the cyclic composition properties of the Pauli matrices, we find
\begin{align}\label{Eq:Ptwirl2rdot}
\frac{1}{2}\Trace{\PTwirl{j}\PTwirl{k}} &= \rhat_{j}\cdot\rhat_{k}.
\end{align}
From Eqs. \ref{Eq:DominantXiTerm}, \ref{Eq:DominantXiOperatorsInTermsOfPtwirls} and \ref{Eq:Ptwirl2rdot} we therefore obtain 
\begin{align}
\HTRBTaylorTerm{2}{1,1} &= 
-\sum_{j<k}\delta_j\delta_k
\rhat_{j}\cdot\rhat_{k}.
\end{align}
Observing the quantity $\delta_j\delta_k \rhat_{j}\cdot\rhat_{k}$ is invariant under exchange of indices, we may recast the restricted sum over $j<k$ into an unrestricted sum, picking up a residual term $\HTRBTaylorTerm{2}{2} = -\frac{1}{2}\sum_{j=1}^J\delta_j^2$ (see Appendix \ref{Appendix:ApproximatingFidNoiseAv}), to obtain  
\begin{align}\label{Eq:Q211}
\HTRBTaylorTerm{2}{1,1} =-\frac{1}{2}\norm{\RvecA}^2-\HTRBTaylorTerm{2}{2},\;\;\;\;\;\RvecA\equiv\sum_{j=1}^{J}
\delta_{j}\rhat_{j}.
\end{align}
Consequently, $\RvecA$ is a random walk in ``Pauli space''.

Taking half the trace of Eq. \ref{Eq:RBNoisySequenceTaylorExpandExpression2} and substituting in Eq. \ref{Eq:Q211}, the term $\HTRBTaylorTerm{2}{2}$  cancels out in the expression for $\frac{1}{2}\text{Tr}\left(\RBseqNoisy_{\CliffVec,\NoiseVec}\right)$. This further simplifies by observing that
\begin{align}\label{Eq:UsefulProperty1}
\frac{1}{2}\Trace{
\Crun{1,j-1}\Z^{k}\Crun{j,J}}= \begin{cases}
0,\hspace{0.75cm}&k\hspace{0.5cm}\text{odd}\\
1,\hspace{0.75cm}&k\hspace{0.5cm}\text{even}
\end{cases}
\end{align}
This follows from (a) the cyclic property of the trace, (b) the technical constraint $\RBseqClean_{\CliffVec} \equiv\Crun{1,J} \equiv \Identity$, 
 and (c) $\Z^k$ is either $Z$ or $\Identity$ depending on whether $k$ is odd or even respectively. 
 Using Eq. \ref{Eq:UsefulProperty1} we find $\HTRBTaylorTerm{0}{} =1$ and $\HTRBTaylorTerm{1}{1}  =\HTRBTaylorTerm{3}{1,2} = \HTRBTaylorTerm{3}{3}=0$, leading to the simplified expression 
\begin{align}
\label{Eq:HTRBseqNoisyExpansion}
\frac{1}{2}\Trace{\RBseqNoisy_{\CliffVec,\NoiseVec}}&\approx 
1
-\frac{1}{2}\norm{\RvecA}^2
+\HTRBTaylorTerm{3}{1,1,1} \\
&+\HTRBTaylorTerm{4}{1,1,1,1} 
+\HTRBTaylorTerm{4}{1,1,2} 
+\HTRBTaylorTerm{4}{1,3} 
+\HTRBTaylorTerm{4}{2,2} 
+\HTRBTaylorTerm{4}{4} \nonumber.
\end{align}
Substituting this into Eq.\ \ref{Eq:NoiseAveragedFidelityDefinition} and retaining only terms up to $\Order{\sigma^4}$ we obtain 
\begin{align}
\label{Eq:NAFExpansion1}
\FidNoiseAv&\approx 
1
-\langle\norm{\RvecA}^2\rangle
+O^{(4)},
\end{align}
where 
\begin{align}\label{Eq:4thOrderResidualTerm}
O^{(4)} &\equiv \frac{1}{4}\left\langle{\norm{\RvecA}^4}\right\rangle
+2\left\langle\HTRBTaylorTerm{4}{2,2}\right\rangle
+2\left\langle\HTRBTaylorTerm{4}{4} \right\rangle.
\end{align}
To a good approximation $O^{(4)}$ may be treated as a small correction in the form of a small constant, with the statistical distribution properties residing in the leading-order term, the random variable $\langle\norm{\RvecA}^2\rangle$.   In arriving at these expressions we have used the assumption that the error process has zero mean, from which it follows only terms with noise random variables raised to even powers, or those summed over terms raised to even powers, survive the ensemble average with the others reducing to zero. In particular, 
$\langle\HTRBTaylorTerm{3}{1,1,1} \rangle
= \langle\HTRBTaylorTerm{4}{1,1,1,1} \rangle
=\langle\HTRBTaylorTerm{4}{1,1,2} \rangle
=\langle\HTRBTaylorTerm{4}{1,3} \rangle
=0$ (see Appendix \ref{Appendix:ApproximatingFidNoiseAv}).


\vspace{\baselineskip}
\section{Probability Distribution Functions for RB Outcomes via a Geometric Interpretation of Error Accumulation}
In the expressions above we see that the key metric capturing the reduction of fidelity in a randomized benchmarking sequence is
\begin{align}\label{Eq:DefinitionRvec}
\RvecA\equiv\sum_{j=1}^{J}
\delta_{j}\rhat_{j}
\end{align}
which may be interpreted as a random-walk in $\mathbb{R}^3$ ``Pauli space'' generated by adding $J$ randomly-oriented steps along the principle Cartesian axes, with step lengths specified by $\NoiseVec$.  Walks which terminate far from the origin correspond to sequences with large net infidelities while those which ultimately end near the origin have small infidelities.  These observations form a key contribution of this work, as we will show. 

This geometric picture provides a unique insight into how error accumulates in RB sequences, and facilitates calculation of the distribution of $\FidNoiseAv$ incorporating the effects of correlations in the error process.  In particular, the calculation of sequence fidelities maps onto the distribution of random walk, with possible correlations manifesting the random step lengths. 

  In the following sections we explore how correlations in $\NoiseVec$ affect the distributions of the terms in Eq. \ref{Eq:NAFExpansion1}, and hence the probability density function (PDF) $f_{\FidNoiseAv}(F)$ of the noise-averaged fidelity $\FidNoiseAv$.   Our calculations demonstrate that under all error models treated here $f_{\FidNoiseAv}(F)$ takes the form of a gamma distribution, which is well known in statistics and provides the significant benefit in that explicit analytic forms are available for both the moments of the distribution and the moment-generating functions (see Table~\ref{Table:T1}). Interestingly the gamma distribution is known to be useful for failure analysis and life-testing in engineering~\cite{Papoulis} which bears some similarity to the notion of error accumulation in RB.  


\subsection{PDF for Markovian Processes}\label{Sec:Markovian}
				
%
In the Markovian limit (\emph{i.e.}\ uncorrelated noise), we assume all noise random variables $\delta_j$ are i.i.d. Hence $\RvecA$ corresponds to a $J$-length unbiased random walk with step lengths sampled from the normal distribution $\mathcal{N}\left(0,\hspace{0.05cm}\sigma^2\right)$. Since these step lengths are symmetrically distributed about zero, the distributions of the components of the walk vector $\delta_j\rhat_{j} = (\delta_jx_{j},\hspace{0.05cm}\delta_jy_{j},\hspace{0.05cm}\delta_jz_{j})$ are invariant with respect to the sign of the coefficients $\alpha_{j}$ in all Cartesian directions $\alpha \in\{x,y,z\}$. Ignoring the signs we therefore treat the the coefficients as binaries $\alpha_j\in\{0,1\}$, where the zero event simply reduces the number of steps taken in that direction. Let 
\begin{align}
\hspace{-0.25cm}
n_{\alpha} \equiv \sum_{i=1}^J\abs{\alpha_{j}},
\hspace{0.4cm}
\alpha\in\{x,y,z\},
\hspace{0.4cm}
n_x+n_y+n_z = J
\end{align}
count the total number of nonzero components in each Cartesian direction over the sequence of walk vectors $\{\rhat_{1},\rhat_{2},...,\rhat_{J}\}$. Then
\begin{align}
\hspace{-0.2cm}
\RvecA
&=\left(
\delta^{x}_1+...+\delta^{x}_{n_x},
\hspace{0.2cm}
\delta^{y}_1+...+\delta^{y}_{n_y},
\hspace{0.2cm}
\delta^{z}_1+...+\delta^{z}_{n_z}
\right)
\end{align}
where the superscripts in $\delta^{\alpha}_j$ indicate summing only over the subset of $\delta_j$ for which the coefficients $\alpha_{j}$ are nonzero. Thus we have 
\begin{align}
\norm{\RvecA}^2 &=\Delta_x^2+\Delta_y^2+\Delta_z^2
\end{align}
where we have defined $\Delta_\alpha\equiv
\left(\delta_1^{\alpha}+\delta_2^{\alpha}+...+\delta_{n_\alpha}^{\alpha}\right)$ for $\alpha\in\{x,y,z\}$.
Since all $\delta_j\sim\mathcal{N}\left(0,\hspace{0.05cm}\sigma^2\right)$ are i.i.d., the new random variables $\Delta_{\alpha}\sim\mathcal{N}\left(0,\hspace{0.05cm}n_{\alpha}\sigma^2\right)$ are also independent and normally distributed, with zero mean and variance $n_{\alpha}\sigma^2$.
\indent The distribution of the sum of squares of non-identical Gaussians is generally complicated to write down, requiring a generalized chi-square distribution. To avoid this we make the following modest approximation. Since the vectors $\rhat_{j}$ are uniformly-distributed there is a $\frac{1}{3}$ probability of being parallel to any given Cartesian axis. The probability of finding any particular combination $(n_x,n_y,n_z)$ is therefore given by the multinomial distribution 
\begin{align}
\mathcal{P}\left(n_x,n_y,n_z\right) &= \frac{J!}{n_x!n_y!n_z!}\left(\frac{1}{3}\right)^{n_x}\left(\frac{1}{3}\right)^{n_y}\left(\frac{1}{3}\right)^{n_z}
\end{align}
For $J\gtrsim5$, however, this is sufficiently peaked around $n_{x,y,z} = J/3$ that we may simply regard these values as fixed without significant error. In this case $\Delta_{x,y,z}\sim\mathcal{N}\left(0,\hspace{0.05cm}J\sigma^2/3\right)$ reduce to i.i.d. random variables. The distribution of $\norm{\RvecA}^2$ consequently reduces to chi-square distribution with 3 degrees of freedom. It is more convenient, however, to express this in more general terms as a member of the two-parameter family of \emph{gamma distributions} (see Eq. \ref{AppendixEq:DefinitionGammaDistribution}), of which the chi-square is a special case. Specifically, we obtain 
\begin{align}\label{Eq:RSQGammaDistribution}
&\norm{\RvecA}^2\sim\Gamma\left(\alpha,\beta\right),
\hspace{0.5cm}
\alpha = \frac{3}{2},
\hspace{0.5cm}
\beta=\frac{2J\sigma^2}{3}
\end{align}
with shape parameter $\alpha$ and scale parameter $\beta$. An ensemble average over $n$ independent noise realizations is therefore specified by 
\begin{align}
\langle \norm{\RvecA}^2\rangle_n = \frac{1}{n}\sum_{j=1}^n\norm{\RvecA}_j^2,
\hspace{0.5cm}\norm{\RvecA}_j^2\sim\Gamma\left(\frac{3}{2},\frac{2J\sigma^2}{3}\right)
\end{align}
where the $\norm{\RvecA}_j^2$ are i.i.d. gamma-distributed random variables. But the sample mean over $n$ gamma-distributed random variables simply yields a rescaled gamma distribution with $\alpha\rightarrow n\alpha$ and $\beta\rightarrow\beta/n$ (see Eq. \ref{AppendixEq:SampeMeanGammaDistributedRVs}). Consequently 
\begin{align}\label{Eq:RSQnGammaDistribution}
\langle \norm{\RvecA}^2\rangle_n \sim\Gamma\left(\frac{3n}{2},\frac{2J\sigma^2}{3n}\right)
\end{align}
with expectation $\E{\langle \norm{\RvecA}^2\rangle_n }=J\sigma^2$ and variance $\Var{\langle \norm{\RvecA}^2\rangle_n }= \frac{2}{3}J^2\sigma^4n^{-1}$.

Higher-order contributions may be included by computing the terms in $O^{(4)}$. For full derivations see Appendix \ref{Appendix:MarkovianDerivation}; here we sketch the result. From the known distribution of $\norm{\RvecA}^2$ in Eq \ref{Eq:RSQGammaDistribution}, the PDF for $\norm{\RvecA}^4$ is given by a transformation allowing us to compute the expectation and variance. Taking an ensemble average over noise realizations, applying the central limit theorem, and observing the narrowness of the resulting distribution compared to the leading-order term, we make the approximation 
$\langle\norm{\RvecA}^4\rangle\approx \E{\norm{\RvecA}^4} = \frac{5}{3}J^2\sigma^4$. The remaining terms yield values $\langle\HTRBTaylorTerm{4}{2,2}\rangle=\frac{1}{8}J(J-1)\sigma^4$ and $\langle\HTRBTaylorTerm{4}{4}\rangle  =\frac{1}{8}J\sigma^4$. Substituting these into Eq. \ref{Eq:4thOrderResidualTerm} we obtain the 4th order correction $O^{(4)} = \frac{2}{3}J^2\sigma^4$ and the noise-averaged fidelity reduces to 
\begin{align}
\label{Eq:NAFfinalMK}
\FidNoiseAv&\approx 
1
-\langle\norm{\RvecA}^2\rangle_n
+\frac{2}{3}J^2\sigma^4
\end{align}
inheriting the gamma distribution of $\langle\norm{\RvecA}^2\rangle_n$ described in Eq. \ref{Eq:RSQnGammaDistribution}. We linearly transform this expression to produce a final PDF for noise-averaged \emph{fidelity} in the Markovian regime: 
\begin{equation}
\label{Eq:NAF PDF MK}
f_{\FidNoiseAv}(F) \equiv \nu(F)^{\alpha-1}e^{-{\nu(F)}/{\beta}}\beta^{-\alpha}/\Gamma(\alpha),
\end{equation}
where $\Gamma(x)$is the gamma function, and  the quantities $\nu(F)$, $\alpha$ and $\beta$ are defined in Table~\ref{Table:T1}.



\subsection{PDF for DC (quasi-static) processes}\label{Sec:DC}
				
%
In the DC limit (\emph{i.e.}\ quasi-static noise) we assume all noise random variables $\delta_j\equiv\delta$ are identical (maximally correlated) over a given sequence $\RBseqClean_{\CliffVec}$. However over separate instances $\delta$ is sampled from the normal distribution $\delta\sim\mathcal{N}\left(0,\hspace{0.05cm}\sigma^2\right)$. Then $\RvecA$  corresponds to a $J$-step unbiased random walk with \emph{fixed} step length $\delta$ directed along the Cartesian axes. In this case the noise random variables $\delta$ and Clifford-dependent random variables $\rhat_{j}$ factorize, allowing us to express 
\begin{align}\label{Eq:DefinitionOfVvec}
\RvecA=\delta\Vvec,
\hspace{0.5cm}
\Vvec\equiv\sum_{j=1}^J\rhat_{j}
\end{align}
where $\Vvec\in\mathbb{R}^3$ defines an unbiased random walk on a 3D lattice generated by adding $J$ \emph{unit-length} steps.  Thus, in contrast with the Markovian case, here the random walk in Pauli space is unaffected (step-by-step) by the noise interactions. Rather, in a given run, the noise variables $\delta$ effectively scale the random walk $\Vvec$ generated by the Clifford sequence, up to a sign. Since we are interested in the norm square
$\norm{\RvecA}^2 = \delta^2\norm{\Vvec}^2$, however,
any sign dependence of $\delta$ vanishes. Performing a finite ensemble average over $n$ noise randomizations we therefore obtain 
$
\langle\norm{\RvecA}^2\rangle_n 
= 
\langle\delta^2\rangle_n
\norm{\Vvec}^2$, and Eq. \ref{Eq:NAFExpansion1} yields
\begin{align}\label{Eq:NAFapproxDC}
\FidNoiseAv\approx 
1
-\langle\delta^2\rangle_n\norm{\Vvec}^2
+
O^{(4)}
\end{align}
In this case, $O^{(4)}$, includes a term $\frac{1}{4}\langle \delta^4\rangle_n\norm{\Vvec}^4$ which is now \emph{highly correlated} with the leading-order contribution, so we cannot use the expectation value as a proxy for the whole distribution.  However corrections from these terms are $O(\sigma^4)$, so we ignore these terms, and formally study the limit $J\sigma^2\ll1$.  Numerical evidence indicates that this approximation works well up to $J\sigma^2\sim1$.


Since $\delta\sim \mathcal{N}\left(0,\sigma^2\right)$ is normally distributed, $\delta^2$ is chi-square-distributed which is a special case of the gamma distribution, $\delta^2\sim\Gamma(\alpha,\beta)$ with shape parameter $\alpha = 1/2$ and scale parameter $\beta = 2\sigma^2$. Taking the sample mean over $n$ independent gamma-distributed variables again yields a rescaled gamma distribution with $\alpha\rightarrow n\alpha$ and $\beta\rightarrow\beta/n$ 
\begin{align}
\left\langle\delta^2\right\rangle_n\sim\Gamma\left(\frac{n}{2},\frac{2\sigma^2}{n}\right)
\end{align}
with expectation  
$\E{\langle \delta^2\rangle_n }=\sigma^2$ and variance $\Var{\langle \delta^2\rangle_n }= \frac{2\sigma^4}{n}$.

The random-walk behavior of $\norm{\Vvec}^2$ (Eq. \ref{Eq:DefinitionOfVvec}) represents a well-studied problem in diffusion statistics. Let the random variable $R$ be the distance from the origin in a symmetric (Bernoulli) 3D random walk after $J$ steps. It is straightforward to show that the PDF for $R$ is 
\begin{align}
\label{Eq:BernoulliRandomWalkDistancePDF}
f_{R}(r) = \left(\frac{3}{2\pi J}\right)^{3/2}4\pi r^2 e^{\frac{-3r^2}{2J}}.
\end{align}
This expression describes a random walk of unit step length, and is derived assuming that $\rhat_j$ is uniformly and continuously sampled from \emph{all} directions in $\mathbb{R}^3$, however \eqn{Eq:BernoulliRandomWalkDistancePDF} is a good approximation for the PDF of a walk on a 3D lattice  \cite{Note1}. The distribution of the distance square $\norm{\Vvec}^2 \approx R^2$ is then given by the transformation (see Eq. \ref{AppendixEq:PDFIncreasingFunctionOfRV})
\begin{align}
f_{\norm{\Vvec}^2}(x) &= \frac{1}{2 x^{1/2}}f_{R}\left(x^{-1/2}\right)\\
&= \frac{1}{\Gamma(\alpha)\beta^\alpha}x^{\alpha-1}\exp\left(-\frac{x}{\beta}\right)
\end{align}
where $\alpha = 3/2$ and $\beta = 2J/3$, and $\Gamma(x)$ is the gamma function; again this is a gamma distribution (see Eq. \ref{AppendixEq:DefinitionGammaDistribution}) with shape parameter $\alpha$ and scale parameter $\beta$. Consequently 
\begin{align}\label{Eq:DistributionOfVvec}
\norm{\Vvec}^2\sim\Gamma\left(\frac{3}{2},\hspace{0.05cm}\frac{2J}{3}\right).
\end{align}

Thus, to leading order, the PDF for $\FidNoiseAv$ with DC noise is specified by the product of two independent gamma-distributed random variables. The closed-form expression can be calculated by direct integration (see Appendix \ref{Appendix:DCDerivation}), however for moderate ensemble sizes $n\gtrsim 50$ it is sufficient to approximate $\langle\delta^2\rangle_n$ as strongly peaked around its mean, $\sigma^2$, such that $\FidNoiseAv\approx 1-\sigma^2\norm{\Vvec}^2$.  In this case the PDF $f_{\FidNoiseAv}(F)$ reduces to a linear-transformed gamma distribution associated with $\norm{\Vvec}^2$ and possesses \emph{the same form as} Eq.~\ref{Eq:NAF PDF MK} but with different values of the parameters $\nu(F)$, $\alpha$ and $\beta$, as given defined in Table~\ref{Table:T1}.  This is a remarkable observation given the substantial differences in error correlations between these extreme limits.

\begin{table}[tp]
\footnotesize
\renewcommand{\arraystretch}{1.4}
\begin{tabular}{|l|c|c|c|}\hline
       & Markovian                                           & DC      				&Block-correlated                   \\ \hline\hline 
            
$\alpha$ & $\frac{3}{2}n$                                                           & $\frac{3}{2}$     		&$\frac{3}{2}J/(M-1)$               \\\hline 
            
$\beta$  & $\frac{2}{3}J\sigma^{2}/n$                                   & $\frac{2}{3}J\sigma^{2}$	&$\frac{2}{3}(M-1)\sigma^{2}$			\\\hline 
            
$\nu(F)$    & $1-F+ \frac{2}{3}J^{2}\sigma^{4}$  & $1-F$				&$1-F$	\\           \hline \hline
$\mathbb{E}$ & $1-J\sigma^{2}+\frac{2}{3}J^{2}\sigma^{4}$  & $1-J\sigma^{2}$ 	& $1-J\sigma^{2}$ \\ \hline
$\mathbb{M}$ & $1-J\sigma^{2}\left(1-\frac{2}{3n}\right)+\frac{2}{3}J^{2}\sigma^{4}$ & $1- \frac{1}{3}J\sigma^{2}$ 	&$1- J\sigma^{2}\left(1-\frac{2}{3}\frac{M-1}{J}\right)$\\ \hline
$\mathbb{V}$ &$\frac{2}{3}J^{2}\sigma^{4}/n$&$\frac{2}{3}J^{2}\sigma^{4}$	&$\frac{2}{3}J(M-1)\sigma^{4}$	\\ \hline
$\mathbb{S}$ &$-2\sqrt{2/3n}$ &$-2\sqrt{2/3}$	&$-2\sqrt{2(M-1)/3J}$	\\     \hline
\end{tabular}\normalsize
\caption{Scale and shape parameters for the respective gamma distributions, and calculated moments for noise-averaged fidelity distributions $f_{\FidNoiseAv}(F)$.  Moments obtained from standard gamma distribution after appropriate linear transformation specified by $\nu(F)$: \mbox{Expectation, $\mathbb{E}\left[\FidNoiseAv\right]=\nu+F-\alpha\beta$}; Mode, $\mathbb{M}\left[\FidNoiseAv\right]=\nu+F-(\alpha-1)\beta$; \mbox{Variance, $\mathbb{V}\left[\FidNoiseAv\right]=\alpha\beta^{2}$}; Skew, $\mathbb{S}\left[\FidNoiseAv\right]=-2/\sqrt{\alpha}$.  }
\label{Table:T1}
\end{table}


\subsection{PDF for Generically-Correlated Processes}\label{Sec:PSD}
				
%

For both limiting cases of Markovian and DC noise correlations treated above we showed $\FidNoiseAv$ is described, to first order, by the gamma distribution with shape and scale parameters dependent on the correlation structure. Assuming continuity of the distribution we also expect $\FidNoiseAv$ to be approximately gamma-distributed for generic, intermediate correlation structures.  We can show this formally for a specific class of correlated noise models, in which the noise is \emph{block-correlated}; \emph{i.e.} \ the error random variables $\delta_j$ are constant over blocks, or subsequences, of Clifford operators of fixed length $M\le J$,  and there is no correlation between distinct blocks. The full derivation of the PDF for this case can be found in Appendix \ref{Sec:BlockCorrelated}.  We find block-correlated noise yields a gamma-distributed fidelity, with parameters $\alpha={3J}/{2(M-1)}$, $\beta={2(M-1)\sigma^2}/{3}$, enabling us to interpolate between the Markovian ($ M=1$) and DC ($M=J$) limits for arbitrary correlation length $M$ (see Table \ref{Table:T1}). 

While block-correlated noise is not stationary (\emph{i.e.}\ it does not have a stationary power spectrum),  it simply and explicitly captures the notion of a correlation length, $M$.  The correlation length thus manifests itself in the distribution of fidelities. With these insights, for brevity we assume that  generic noise correlations also give rise to gamma-distributed infidelities.  This is supported by the quantitative calculations and qualitative arguments in Appendix \ref{Sec:BlockCorrelated}, and by comparison with Monte Carlo numerics.  Consequently, the shape and scale factors can be inferred from the mean and variance of the distribution, which we calculate directly.

For simplicity, we assume the error process is sufficiently non-Markovian that the noise ensemble size $n$ and the $\Order{\sigma^4}$ contributions may be ignored without introducing large errors, as in the DC case. Thus, we write
\begin{align}
&\FidNoiseAv\approx 1-\langle\norm{\RvecA}^2\rangle,
\hspace{0.75cm}
\langle\norm{\RvecA}^2\rangle\sim\Gamma\left[\alpha,\beta\right].
\end{align}
From the first two moments of the gamma distribution, the expectation and variance of the random-walk variable are $\E{\langle\norm{\RvecA}^2\rangle} = \alpha\beta$ and $\Var{\langle\norm{\RvecA}^2\rangle} = \alpha\beta^2$, so the shape and scale parameters are given by 
\begin{align}\label{Eq:AlphaAndBetaFromEandV}
\alpha = \frac{\mathbb{E}^2}{\mathbb{V}},
\hspace{1cm}
\beta = \frac{\mathbb{V}}{\mathbb{E}}.
\end{align}
The task of defining the gamma distribution therefore reduces to computing the expectation and variance of $\langle\norm{\RvecA}^2\rangle$, as a function of the generic correlation structure of the error process, $\NoiseVec$. 

An arbitrary time-series $x(t)$ may be characterized by an autocorrelation function $C_{x}(\tau)  = \langle x(t)x(t+\tau)\rangle_t$ where the ensemble average is over all $t$, and $\tau$ is the \emph{time difference} between measurements. In this case, invoking the Wiener-Khintchine theorem, the power spectral-density (PSD) is given by the Fourier transform $S(\omega) = \frac{1}{\sqrt{2\pi}}\int_{-\infty}^{\infty}C_{x}(\tau)e^{i\omega\tau}d\tau$. We make use of these relations by discretizing the time-series. Let the elements of $\NoiseVec$ be defined by $\delta_j=x(t_j)$ for $t_j/\tau_{g} \in \{1,2,..,J\}$ where $\tau_{g}$ is the time taken to perform a Clifford operation. The underlying error process is thereby discretely ``sampled'' by the Clifford sequence $\RBseqClean_{\CliffVec}$, and correlations between elements of $\NoiseVec$ separated by a time interval of ``$k$ gates'' are specified by the (discrete) autocorrelation function
\begin{align}
C_{\NoiseVec}(k) \equiv \left\langle \delta_j \delta_{j+k}\right\rangle_j
\end{align}
where the ensemble average is over the index $j$. Substituting these definitions into Eq. \ref{Eq:DefinitionRvec}  the expectation and variance of $\langle \norm{\RvecA}^2\rangle$ are given by (see Appendix \ref{Sec:STATSforGenericCorrelations})
\begin{align}\label{Eq:StatsForACfunction}
\mathbb{E} = JC_{\NoiseVec}(0),
\hspace{1cm}
\mathbb{V}= \frac{4}{3}\sum_{k=1}^{J-1}\left(J-k\right)(C_{\NoiseVec}(k))^2.
\end{align}
In experiment one typically has access to the PSD $S(\omega)$ rather than the autocorrelation function. However this easily maps to our framework via an inverse Fourier transform
\begin{align}
C_{\NoiseVec}(k) =  \int_{-\infty}^{\infty}S(\omega)e^{-i\omega k\tau_{g}}d\omega.
\end{align}
The PDF therefore has \emph{the same form as} Eq. \ref{Eq:NAF PDF MK} but with $\alpha$ and $\beta$ given by Eq. \ref{Eq:AlphaAndBetaFromEandV}, and $\nu(F) = 1-F$. 

With these expressions we now have a complete analytic representation of the distribution of measured noise-averaged fidelities over different RB sequences. In the next section we verify these results with numeric Monte Carlo simulations and discuss the differences in distribution characteristics depending on noise correlations.


\subsection{Comparing PDFs for Various Noise Correlations}\label{Sec:ComparingPDFs}
%
\begin{figure}[t]
\hspace*{-0.5cm}
\includegraphics[width=1.1\columnwidth]{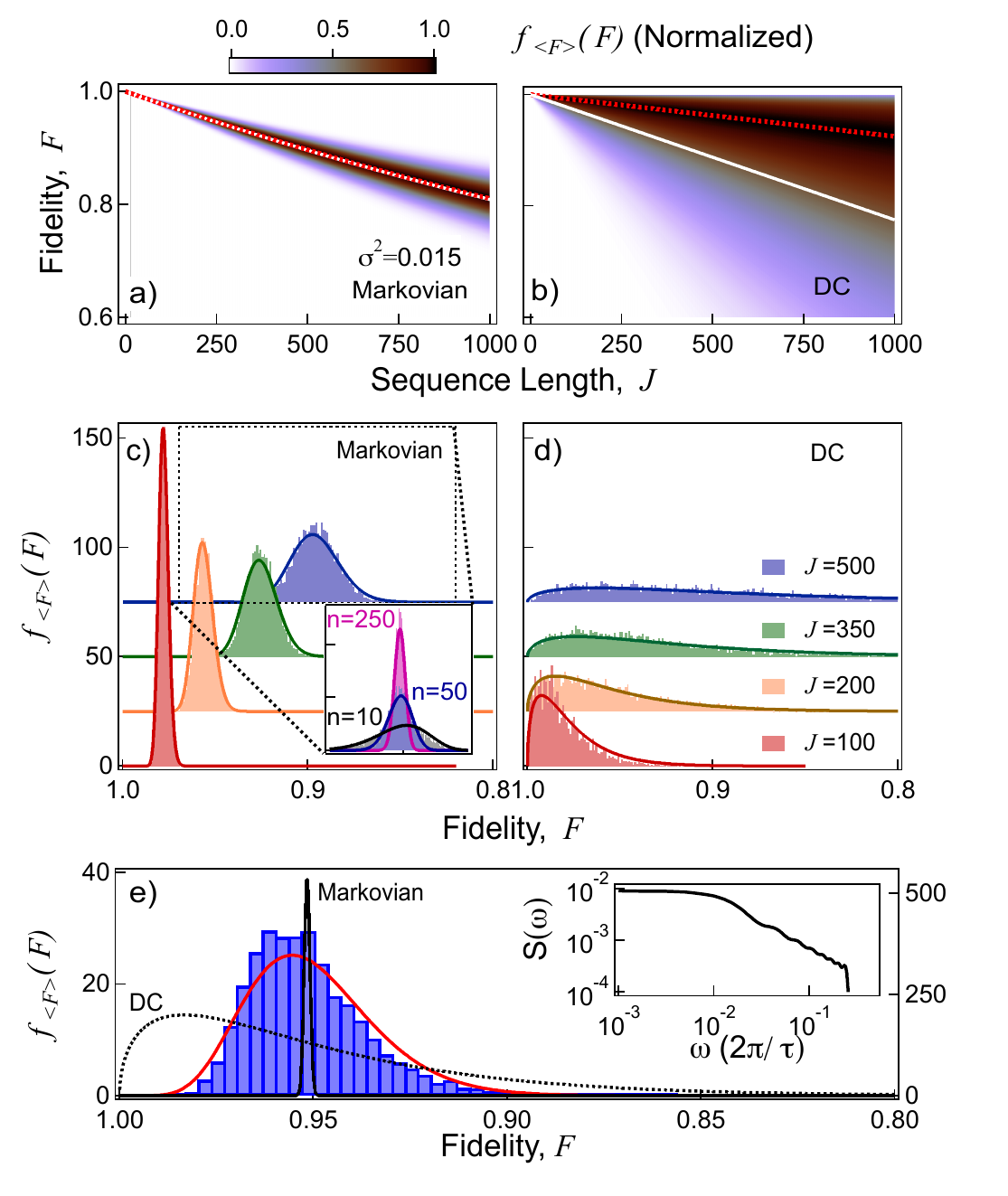}
\vspace{-5mm}
\caption{Analytic PDFs $\FidNoiseAv$ and simulated distributions for different noise regimes using $p_{\text{RB}}\sim2\times10^{-4}$. \textbf{a, b}) $f_{\FidNoiseAv}(F)$ calculated as a function of $J$ for Markovian and DC regimes, normalized to unity at the mode for each $J$ for clarity.  White line indicates analytic mean, $\mathbb{E}[\FidNoiseAv]$; red dashed line indicates analytic mode, $\mathbb{M}[\FidNoiseAv]$.   \textbf{c, d}) Compare analytics (solid lines) to numerically simulated histograms of distributions for various $J$, using $n=50$ and no free parameters.  Curves vertically offset by for clarity by multiples of 25 units.  Inset \textbf{c}) Markovian distributions varying $n$=10, $50$, $250$ (black, blue, pink), corresponding to distribution in dotted box.  \textbf{e}) Analytic PDF $\FidNoiseAv$ (solid red line) and simulated distributions for quasi-$1/\omega$ noise regime with $J = 200$ and $n = 3000$. Comparative PDF for DC (left axis) and Markovian noise (right axis) shown as black dashed and solid lines respectively. Noise parameters chosen such that the mean error is $0.05$. PSD shown in inset, constructed using Fourier synthesis as described in~\cite{SoarePRA2014}.}
\label{Fig:F1}
\end{figure}

The analytic forms for $f_{\FidNoiseAv}(F)$ derived above serve as a tool to analyze the impact of temporal noise correlations on RB experiments. In all cases (Markovian, DC, and Intermediate) the PDF is $\Gamma(\alpha,\beta)$-distributed, with differences captured in the values of the shape parameter $\alpha$ and scale parameter $\beta$. This result is derived from statistics of a 3D random walk. 

Plotting the PDF $f_{\FidNoiseAv}(F)$ in Fig.~\ref{Fig:F1} immediately reveals substantial differences in the distribution of outcomes for the two limiting cases. While the distributions in both Markovian and DC cases yield the same value for the mean (the statistic currently used in RB protocols), the higher order moments diverge significantly. For DC noise $f_{\FidNoiseAv}(F)$ is skewed towards high-fidelities and possesses a variance significantly larger than that for Markovian errors of equivalent strength, parameterized by the value of $\sigma^{2}$. Averaging over a large noise ensemble results in the mode converging to the mean in the Markovian regime, but maintaining a fixed \emph{higher} value of fidelity for the DC regime. By comparison, the variance and skew for Markovian noise diminish with increasing noise averaging, but remain fixed and nonzero in the DC case.  

To compare our analytic PDFs with the true distributions we directly simulate the fidelity outcomes associated with the metric in Eq. \ref{Eq:TraceFidelityDefinition}. For a given $J$, we generate an ensemble $\{\CliffVec_{1},...,\CliffVec_{k}\}$ of $k$ random Clifford sequences. The first $J-1$ elements in each sequence $\CliffVec_{i}$ are uniformly and independently sampled from the set $\{1,...,24\}$, with the final element chosen such that the total operator product performs the identity $\RBseqClean_{\CliffVec_{i}} \equiv \Identity$. For each sequence $\CliffVec_{i}$ we then generate an ensemble $\{\NoiseVec_{i,1},...,\NoiseVec_{i,n}\}$ of $n$ random and independent realizations of the error process, where each $\NoiseVec_{i,j}$ is a sequence of $J$ noise random variables generated by Monte Carlo sampling from the appropriate correlated-error model. For Markovian and DC processes, this is fully described in Section \ref{Sec:PhysicalModel}. For the intermediate case, random sequences $\NoiseVec$ with a desired PSD $S(\omega)$ and autocorrelation function $C_{\NoiseVec}(k)$ may be generated by uniformly-phase-randomized Fourier synthesis as described in~\cite{SoarePRA2014} (see Appendix \ref{Appendix:FourierSynthNoise}).

For each pair $\CliffVec_{i}$ and $\NoiseVec_{i,j}$ the operator product in Eq. \ref{Eq:RBNoisySequenceFullExpression} is computed, using Table \ref{Table:CliffordGroup} to determine the ($2\times2$) unitary matrix representing each Clifford operator. For each pair the trace fidelity  $\Fid{i,j} =  \mathcal{F}(\CliffVec_{i}, \NoiseVec_{i,j})$ is then calculated using Eq. \ref{Eq:TraceFidelityDefinition}, generating the array shown in  Eq. \ref{Eq:FiniteRBSampleSpaceArray}, simulating the measured fidelity outcomes.  Averaging over columns, as in Eq. \ref{Eq:SimulatedNAF}, the array reduces to a column vector containing $k$ noise-averaged fidelities $\overline{\boldsymbol{F}}^{(J)}_{i,\langle\cdot\rangle}$, one for each $\CliffVec_{i}$, which we finally plot as a normalized histogram and compare against our analytic PDFs.

Fig.~\ref{Fig:F1}(c) and (d) compares the numerically generated histograms of fidelity against the analytic results,  Eq. \ref{Eq:NAF PDF MK}, for Markovian and DC noise respectively.  These are are in excellent agreement  for the different error processes considered. The characteristic long-tailed distribution peaked near high fidelities in the DC limit reproduces key features observed in recent experiments~\cite{BrownPRA2011}. Agreement with analytics is good (with no free parameters) for $J\sigma^2\lesssim1$, beyond which higher order error terms contribute to the distribution.  

For correlated noise, similarly good agreement is obtained.  We have validated the block-correlated noise model (not shown).   For quasi-$1/\omega$ noise, the fidelities are also well described by the gamma distribution, accounting for correlation length, as shown in \mbox{Fig.\ \ref{Fig:F1}(e)}.  The dominance of low-frequency components in this PSD (see inset) skews $f_{\FidNoiseAv}(F)$ toward higher fidelities than the mean, but the presence of higher-frequency components (shorter correlation times) reduces the variance and partially restores symmetry.  In the limit of power spectra containing substantial high-frequency noise, small shifts due to higher-order terms in Eq. \ref{Eq:4thOrderResidualTerm} become important.

\begin{figure}[t]
\centering
\hspace*{-0.25cm}
\includegraphics[width=1.05\columnwidth]{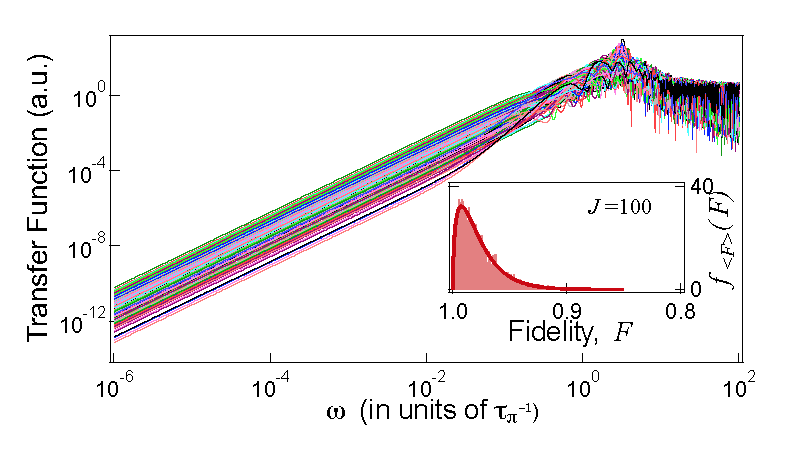}
\vspace{-5mm}
\caption{Filter functions $\{G(\omega; \CliffVec)\}$ for ensemble of RB sequences. Dimensionless angular frequency $\omega$ normalized to the duration of a bit-flip operation, $\tau_{\pi}$; low-frequency ($\omega < 1$) noise susceptibility is captured by vertical offsets and slopes. Inset: Histogram of calculated fidelities for $\{G(\omega; \CliffVec)\}$ using $S(\omega)\propto\delta(\omega-4\pi\times10^{-4}\tau^{-1}_{\pi})$ for quasi-DC noise (scaled to correspond to $\sigma=0.015$) and overlaid with $f_{\FidNoiseAv}(F)$.}
\label{Fig:F2}
\end{figure}

The underlying physical reason for the differences in the two limiting error models is revealed by examining the filter-transfer functions $G(\omega;\CliffVec)$ for the various Clifford randomizations~\cite{GreenNJP2013}.  The noise-averaged fidelity is given by \mbox{$\langle \mathcal{F}\rangle_{\infty}\approx\big(1-e^{-\frac{1}{2\pi}\int_{0}^{\infty}G(\omega;\CliffVec)S(\omega)\omega^{-2}d\omega}\big)/2$}, quantifying the susceptibility to error over a given frequency band, and has demonstrated experimental applications~\cite{SoareNatPhys2014}.  Using techniques outlined in Refs.~\cite{GreenPRL2012, GreenNJP2013} we calculate $G(\omega;\CliffVec)$ for $10^3$ random sequences $\CliffVec$, shown in Fig.~\ref{Fig:F2}. In the low-frequency regime we observe variations over several orders-of-magnitude in the vertical offset of $G(\omega; \CliffVec)$ and also variations in slope at higher frequencies, indicative of partial error cancellation and hence substantial variations in susceptibility to correlated errors. The corresponding distribution of fidelities agrees well with $f_{\FidNoiseAv}(F)$ for DC noise (see inset). 

These observations arise from the fact that some  RB sequences contain coherent, error-suppressing  subsequences.  
In fact RB sequences bear resemblance~\cite{Laflamme2009} to randomized dynamical decoupling protocols known to suppress errors in certain limits of correlated noise~\cite{ViolaRDD}. This would lead to ``artificially'' small measured error for correlated noise, thereby increasing the variance and skew in $f_{\FidNoiseAv}(F)$ towards high fidelities for sufficiently low-frequency-dominated noise. Furthermore, this explicit link between the form of $f_{\FidNoiseAv}(F)$ and underlying symmetries in $\RBseqClean_{\CliffVec}$ suggests we may downselect RB sequences using, \emph{e.g.} the filter function or appropriate entropy measures on the Clifford sequences to ensure coherent averaging properties are minimized.  To the best of our knowledge this is the first quantitative validation of  the  mechanism underlying the shifts in $f_{\FidNoiseAv}(F)$ for temporally-correlated errors, and the first application of the filter-transfer function formalism for quantum control  \emph{at the algorithmic level}.

\vspace{\baselineskip}
\section{Discussion and Conclusion}\label{Sec:Discussion}

Our primary observation is that the form of  $f_{\FidNoiseAv}(F)$ -- and in particular the moments of the distribution -- can exhibit strikingly different behaviors in different error regimes.  This is true despite the fact that the expectation $\mathbb{E}[\FidNoiseAv]\equiv\mathbb{E}[\mathcal{F}]_{J}$ approximately converges for Markovian and DC cases for weak noise (see Table \ref{Table:T1}).  This has a variety of important impacts in the application and interpretation of RB experiments for quantum information.  

For instance, due to the form of $f_{\FidNoiseAv}(F)$ for low frequency noise, the sample estimator $\hat{\mu}^{(J)}$ obtained from an ensemble of $k\ll 24^{J}$ instances of $\CliffVec$ can differ from the true expectation $\mathbb{E}[\mathcal{F}]_J$, generally leading to \emph{overestimation} of the mean fidelity~\cite{Flammia2014, Epstein2014}. This risks systematically underestimating $p_{\text{RB}}$ due to insufficient sampling over $\CliffVec$.   The difference $|\hat{\mu}^{(J)}-\mathbb{E}[\mathcal{F}]_{J}|$ may be formally bounded using moment-generating functions for the gamma distribution (see \emph{Supplemental Material}) to provide a more inclusive bound than previous approximations~\cite{Flammia2014, Epstein2014}.  Ensuring $\hat{\mu}^{(J)}$ falls within an acceptable confidence interval, say within $\pm10\%$ of $\mathbb{E}[\mathcal{F}]_{J}$ (and assuming \mbox{$\sigma=0.015$}, or $p^{(M)}_{\text{RB}}\sim2\times10^{-4}$) requires $k^{(M)}_{\text{min}}>9$ randomizations be selected in the Markovian case, but $k^{(DC)}_{\text{min}}>443$ for DC case. These values increase rapidly as confidence bounds tighten.   These sample sizes are much larger than typically employed in experimental settings, but may be partially relaxed when calculating $p_{\text{RB}}$ by fitting to measurements performed for multiple values of $J$.  In the Markovian regime we also find a tradeoff between finite noise sampling (experimentally achieved by repeating a fixed sequence $\RBseqClean_{\CliffVec}$ and averaging over the resulting projective measurements), and the required $k^{(M)}_{\text{min}}$.  Increasing the noise averaging, $n$, reduces the value $k^{(M)}_{\text{min}}$.

Beyond the question of how well measurements performed in strongly correlated environments \emph{estimate} $p_{\text{RB}}$, there is uncertainty surrounding the breadth of applicability of this single proxy metric as a general quantum verification tool for quantum information.   One key observation is that while the variance and skew of the $f_{\FidNoiseAv}(F)$ converge to zero for Markovian errors in the limit of infinite noise averaging ($n\to\infty$), both remain fixed for DC noise.  Accordingly our results provide direct evidence of the divergence between $p_{\text{RB}}$ and parameters relevant to fault-tolerance~\cite{Gutierrez2015} such as worst-case errors~\cite{Sanders2015} in noise environments with strong temporal correlations; in the DC limit the worst-case error can be much larger than the average error.  

Finally, the fact that some RB sequences are intrinsically ``blind'' to correlated errors, highlights potential shortcomings in performing experimental gate optimization by maximizing $\hat{\mu}^{(J)}$ at fixed $J$;  an operator may optimize an experimental parameter (to maximize the RB fidelity) in such a way to \emph{increase} systematic errors in individual gates.  Such issues may be partially mitigated by selecting subsets of valid RB sequences using the length of the random walk, $||\RvecA||$, in the DC limit to ensure fidelities are not overestimated in a RB procedure.  Future work will explore the use of entropic measures to associate RB sequence structure with $||\RvecA||$ without the need to perform full calculations of the random walk.

We conclude that the interpretation and applicability of a measured RB outcome $p_{\text{RB}}$ can differ significantly depending on the nature of the underlying errors and measurement parameters experienced in a real experimental situation.   This challenge can be partially mitigated through presentation of more complete datasets -- specifically $\overline{\boldsymbol{F}}_{i,\langle\cdot\rangle}$ for each sequence -- in order to assist readers making comparisons between reported results.  We believe the new insights our calculations have revealed will help bound the utility of $p_{\text{RB}}$ in quantum information settings and also help experimentalists ensure that measurements are not subject to hidden biases.



\begin{acknowledgments}
\textit{Acknowledgements:}  The authors thank J. M. Chow, J. Gambetta, and A. Yacoby for useful discussions on RB measurements, and L. Viola for discussions on randomized decoupling. Work partially supported by the ARC Centre of Excellence for Engineered Quantum Systems CE110001013, the Intelligence Advanced Research Projects Activity (IARPA) through the ARO, the US Army Research Office under Contracts W911NF-12-R-0012, W911NF-14-1-0098 and W911NF-14-1-0103, and a private grant from H. \& A. Harley.  STF acknowledges support from an ARC Future Fellowship FT130101744.
\end{acknowledgments}

\vspace{-6mm}

\bibliography{RB_PRA}

\newpage
\pagebreak
\appendix
\begin{widetext}

\bla
\newpage
\section{Mathematical Preliminaries}\label{Appendix:TransformingAndCombiningRandomVariables}
\bla



\subsection{Linear transformation}
\noindent Let $Y$ is a continuous, non-negative random variable described by probability density functions $f_Y(y)$. If $Z$ is the random variable defined by $Z=\alpha Y + \beta$, where $\alpha,\beta\in\mathbb{R}$ and $\alpha\ne0$, then the probability density functions of $Z$ is 
\begin{align}\label{Eq:Ch6PDFofLTofRV}
f_Z(z) =  \frac{1}{\abs{\alpha}}f_Y\left(\frac{z-\beta}{\alpha}\right)
\end{align}
\subsection{Product distribution}
\noindent Let $X$ and $Y$ be to independent, continuous random variables described by probability density functions $f_X(x)$ and $f_Y(y)$. Then the joint distribution of $Z=XY$ is
\begin{align}\label{AppendixEq:ProductDistribution}
f_Z(z) =  \int_{-\infty}^{\infty}\frac{1}{\abs{x}}f_{X}(x)f_{Y}(z/x)dx 
\end{align}
\subsection{Strictly increasing function of a random variable}
\noindent Let $X$ be an absolutely continuous non-negative random variable described by probability density function $f_X(x)$. Define the transformed random variable $Y = g(X)$ where $g(x)$ is a strictly increasing function. That is $x_1>x_2\iff g(x_1)>g(x_2)$, so that $g(x)$ has a well-defined inverse $g^{-1}(x)$. Then the probability density functions of $Y$ is
\begin{align}\label{AppendixEq:PDFIncreasingFunctionOfRV}
f_Y(y) = f_X(g^{-1}(y))\frac{d g^{-1}(y)}{dy}
\end{align}
\subsection{Sum of continuous random variables}
\noindent Let $X$ and $Y$ be two independent, continuous random variables described by probability density functions$ f_X(x)$ and $f_Y(y)$. Then the probability density functions of the sum $Z = X +Y$ is
\begin{align}
\label{Eq:SumOfIndependentRVs}
f_Z(z) = \int_{-\infty}^{\infty}f_X(z-y)f_Y(y) dy = \int_{-\infty}^{\infty}f_Y(z-x)f_X(x) dx 
\end{align}
\subsection{Gamma distribution}
\noindent The gamma distribution describes a family of continuous probability distributions, related to beta distribution, and arising naturally in processes for which the waiting times between Poisson-distributed events are relevant. The common exponential distribution and chi-squared distribution are special cases. The gamma distribution is a two-parameter distribution, and there are a few different parametrizations in common use. Throughout this paper we parametrize the distribution in terms of its \emph{shape parameter} $\alpha$ and a \emph{scale parameter} $\beta$. Let $X\sim\Gamma\left(\alpha,\hspace{0.05cm}\beta\right)$ be a gamma-distributed random variable, then the probability density function is defined by 
\begin{align}\label{AppendixEq:DefinitionGammaDistribution}
f_{X}(x)=\frac{1}{\Gamma(\alpha)\beta^\alpha}x^{\alpha-1}\exp\left(-\frac{x}{\beta}\right)
\end{align}
where $\Gamma(x)$ is the gamma function. Table \ref{Table:GammaStatistics} contains some relevant statistics.
\begin{table}[H]~\label{Table:GammaStatistics}
\hspace{7.25cm}\begin{tabular}{|c|c|}\hline
mean & $\alpha\beta$                                                                       \\\hline 
variance & $\alpha\beta^{2}$                                                                  \\\hline 
skew &  $2/\sqrt{\alpha}$                                                                     \\\hline 
mode & $(\alpha-1)\beta$                                                                          \\\hline 
\end{tabular}
\caption{Statistics of the gamma distribution $\Gamma(\alpha,\beta)$.}
\end{table}

\subsection{Distribution of sample mean of gamma-distributed random variables}
\noindent Let $X_i\sim\Gamma(\alpha,\beta)$, $i\in\{1,...,n\}$, be a set of $n$ independent and identically distributed random variables sampled from the gamma distribution 
with shape parameter $\alpha$ and scale parameter $\beta$.
Then the sum $Z=\sum_{i=1}^{n}X_i$ follows the transformed gamma distribution 
\begin{align}\label{AppendixEq:SumOfIndependentGammas}
Z\sim
\Gamma(n\alpha,\beta)
\end{align}
The ensemble average $Z/n$ has probability density function given by transforming $nf_{Z}(nx)$. Substituting into Eq. \ref{AppendixEq:DefinitionGammaDistribution}, and defining rescaled parameters  $\alpha\rightarrow n\alpha$ and $\beta\rightarrow\beta/n$, the distribution of the ensemble average $\langle X_i\rangle_n$ is found to be 
\begin{align}
\label{AppendixEq:SampeMeanGammaDistributedRVs}
\langle X_i\rangle_n\sim\
\Gamma\left(n\alpha,\beta/n\right)
\end{align}
\subsection{Lindeberg-Levy central limit theorem}
\noindent Let $\{X_1,X_2,...,X_n\}$ be a set of i.i.d. random variables with expectation $\E{X_i} = \mu$ and variance $\Var{X_i}=\rho^2<\infty$. Then the sample average $S_n=\frac{1}{n}\sum_{i=1}^nX_i$ converges to the normal distribution 
\begin{align}
\lim_{n\rightarrow\infty}\sqrt{n}(S_n-\mu)\sim\mathcal{N}\left(0,\hspace{0.05cm}\rho^2\right)
\end{align}

\bla	
\newpage					
\section{Clifford Group Representation for Single Qubit}\label{Sec:CliffordGroupRepresentation}
\bla
\noindent A unitary operation $\Clifford$ is an element of the \emph{Clifford group} if
\begin{align}\label{Eq:CliffordPauliNormalizer}
\Clifford\PauliGroup\Clifford^\dagger = \PauliGroup
\end{align}
where we have defined the \emph{Pauli group} 
\begin{align}
\PauliGroup = \{\pm \Identity, \pm \X, \pm \Y, \pm \Z\}.
\end{align}
That is, the Clifford group is the normalizer of the Pauli group, where for every Pauli operation $P\in\PauliGroup$, there is another $P'\in\PauliGroup$ such that $\Clifford P\Clifford^\dagger = P'$. For a single qubit, the set of all such $\Clifford$ may be thought of as rotations of the Bloch sphere that permute the orientation of $\pm \X, \pm \Y, \pm \Z$ in the Cartesian basis associated with the Pauli matrices, which we refer to as ``Pauli space.'' To obtain a clearer physical picture of these operations, consider associating $\X$ to any of the six Cartesian axes $\{\pm\xhat,\pm\yhat,\pm\zhat\}$. With this axis fixed, we may rotate the axes about $\X$ into four possible orientations while preserving $\X\Y\Z$ right-handedness. This is the action of the $\Clifford$ group: the group of rotational symmetries of the cube.

We construct our representation as follows. Let $R_i(\theta)$ represent one of nine elementary unitaries generating a \emph{clockwise} rotation (looking down the axis of rotation toward the origin) through angle $\theta \in\{\pi,\pm\pi/2\}$ about axis $i\in\{x,y,z\}$. The three $\pi$ rotations correspond to 
\begin{align}
R_{x,y,z}(\pi) \equiv \X,\Y,\Z
\end{align}
and we use the shorthand 
\begin{align}
R_i^{\pm} \equiv R_i(\pm \pi/2),\hspace{1cm} i\in\{x,y,z\}
\end{align}
for the remaining six $\pi/2$ rotations. For example, the action of the operators $R_i^{\pm}$ on the Pauli operators/axes is 
\begin{align}
&\RxPlus:\hspace{0.2cm}(\X,\Y,\Z)\rightarrow(\X,-\Z,\Y)\\
&\RyPlus:\hspace{0.2cm}(\X,\Y,\Z)\rightarrow(\Z,\Y,-\X)\\
&\RzPlus:\hspace{0.2cm}(\X,\Y,\Z)\rightarrow(-\Y,\X,\Z)
\end{align}
Products of these nine elementary operations generate a representation of the 24 elements of the single-qubit Clifford group as tabulated in Table \ref{Table:CliffordGroup}. We use this prescription to generate numerical simulations verifying our analytic calculations above.

\begin{table}[]
\begin{centering}
      \begin{tabular}{|c||c|c|c|c|c|}
  \hline
   \# & Gate Name & Action on ($\X,\Y,\Z$) & Minimal Sequence(s) & Notes 	 \\ \hline\hline
 $\Clifford_1$ 	& $\Identity$  		& 	$(\X,\Y,\Z)$ 		& 	$\X_i^2$,\hspace{0.3cm}$R_i^{+}R_i^{-}$,\hspace{0.3cm}$i\in\{1,2,3\}$ &identity \\ \hline
 $\Clifford_2$ 	& $\X$ 		& 	$(\X,-\Y,-\Z)$ 	& 	$\X$ 				& 			\\ 
 $\Clifford_3$ 	& $\Y$ 		& 	$(-\X,\Y,-\Z)$ 	& 	$\Y$ 				&$\pi$ rotation	\\ 
 $\Clifford_4$ 	& $\Z$ 		& 	$(-\X,-\Y,\Z)$ 	& 	$\Z$ 				&			\\ \hline
 $\Clifford_5$ 	& $\RxPlus$ 	& 	$(\X,-\Z,\Y)$ 		& 	$\RxPlus$ 			&			\\
 $\Clifford_6$ 	& $\RyPlus$ 	& 	$(\Z,\Y,-\X)$ 		& 	$\RyPlus$ 			&$+\pi/2$ rotations	\\ 
 $\Clifford_7$ 	& $\RzPlus$ 	& 	$(-\Y,\X,\Z)$ 		& 	$\RzPlus$ 			&	\\ \hline
 $\Clifford_8$ 	& $\RxMin$ 	& 	$(\X,\Z,-\Y)$ 		& 	$\RxMin$ 			&	\\ 
 $\Clifford_9$ 	& $\RyMin$ 	& 	$(-\Z,\Y,\X)$ 		& 	$\RyMin$ 			&$-\pi/2$ rotations	\\ 
$\Clifford_{10}$ 	& $\RzMin$ 	& 	$(\Y,-\X,\Z)$ 		& 	$\RzMin$ 			&	\\ \hline
$\Clifford_{11}$ 	& $$ 			& 	$(-\X,-\Z,-\Y)$ 	& 	$\Z\RxPlus$ 		&	\\ 
$\Clifford_{12}$ 	& $$ 			& 	$(-\X,\Z,\Y)$ 		& 	$\Z\RxMin$ 		&	\\ 
$\Clifford_ {13}$ & $$ 			& 	$(-\Y,-\X,-\Z)$ 	& 	$\RzPlus \X$ 		&	\\ 
 $\Clifford_{14}$ & $$ 			& 	$(\Y,\X,-\Z)$ 		& 	$\RzMin \X$ 		&	\\
 $\Clifford_{15}$ & $$			& 	$(-\Y,-\Z,\X)$ 	& 	$\RzPlus\RxPlus$ 	&	\\ 
 $\Clifford_{16}$ & $$ 			& 	$(-\Y,\Z,-\X)$ 	& 	$\RzPlus\RxMin$ 	&	\\ 
 $\Clifford_{17}$ & $$ 			& 	$(-\Z,-\X,\Y)$ 	& 	$\RxPlus\RzMin$ 	&	\\ 
 $\Clifford_{18}$ & $$ 			& 	$(-\Z,-\Y,-\X)$ 	& 	$\Z\RyMin$ 		&	\\ 
 $\Clifford_{19}$ & $$ 			& 	$(-\Z,\X,-\Y)$ 	& 	$\RzPlus\RyMin$ 	&	\\ 
 $\Clifford_{20}$ & $$ 			& 	$(\Z,-\X,-\Y)$ 	& 	$\RzMin\RyPlus$ 	&	\\ 
 $\Clifford_{21}$ & $H$ 		& 	$(\Z,-\Y,\X)$ 		& 	$\Z\RyPlus$ 		&Hadamard	\\ 
 $\Clifford_{22}$ & $$ 			& 	$(\Y,-\Z,-\X)$ 	& 	$\RzMin\RxPlus$ 	&	\\ 
 $\Clifford_{23}$ & $$ 			& 	$(\Z,\X,\Y)$ 		& 	$\RzPlus\RyPlus$ 	&	\\ 
 $\Clifford_{24}$ & $$ 			& 	$(\Y,\Z,\X)$ 		& 	$\RzMin\RxMin$ 	&	\\ \hline
    \end{tabular}\caption{Representation of the Clifford group for a single qubit from products of elementary rotations. Relevant transformations of the coordinate system $\X,\Y,\Z$ under the action of each Clifford shown in column 3. Minimal sequence of elementary operations needed to generate each Clifford shown in column 4. We also indicate how these geometric rotations map to logical operations of interest for quantum information where relevant.}\label{Table:CliffordGroup}
\end{centering}
\end{table}

\bla
\newpage
\section{Approximating the Noise-Averaged Fidelity $\FidNoiseAv$ }\label{Appendix:ApproximatingFidNoiseAv}
\bla

\noindent
Here we present the full derivation of the approximation to the noise-averaged fidelity $\FidNoiseAv$  given in Eq. \ref{Eq:NAFExpansion1} of the main text. Let $\RBTaylorTerm{n}{k_1,k_2,...,k_m}$ denote the 
$\Order{
\prod_{\rho=1}^m
\delta_{j_\rho}^{k_\rho}
} = \Order{\sigma^{n}}$ 
term in the expansion of Eq. \ref{Eq:RBNoisySequenceTaylorExpandExpression1} due to cross-multiplying terms like $(\delta_{j_1}\Z)^{k_1}(\delta_{j_2}\Z)^{k_2}... (\delta_{j_m}\Z)^{k_m}$, where $\sum_{\rho=1}^{m}k_\rho = n$. Retaining only terms up to fourth order ($n=4$), consistent with our Talyor approximation of the error unitaries, we thereby obtain 
\begin{align}
\label{AppendixEq:RBNoisySequenceTaylorExpandExpression2}
\RBseqNoisy_{\CliffVec,\NoiseVec}&\approx 
\RBTaylorTerm{0}{} 
+\RBTaylorTerm{1}{1} 
+\RBTaylorTerm{2}{1,1} 
+\RBTaylorTerm{2}{2} 
+\RBTaylorTerm{3}{1,1,1} 
+\RBTaylorTerm{3}{2,1} 
+\RBTaylorTerm{3}{3} 
+\RBTaylorTerm{4}{1,1,1,1} 
+\RBTaylorTerm{4}{1,1,2} 
+\RBTaylorTerm{4}{2,2} 
+\RBTaylorTerm{4}{4} 
\end{align}
where
\begin{align}
\RBTaylorTerm{0}{} &=\Crun{1,J} \\
\RBTaylorTerm{1}{1} & =\sum_{j=1}^{J}(i\delta_j)\Crun{1,j-1}\Z\Crun{j,J}\\
\RBTaylorTerm{2}{1,1}  &=\sum_{j<k}(i\delta_j)(i\delta_k)\Crun{1,j-1}\Z\Crun{j,k-1}\Z\Crun{k,J}\\
\RBTaylorTerm{2}{2}  &=-\frac{1}{2}\sum_{j=1}^J\delta_j^2\Crun{1,j-1}\Z^2\Crun{j,J}\\
\RBTaylorTerm{3}{1,1,1} & =\sum_{j<k<l}(i\delta_j)(i\delta_k)(i\delta_l)\Crun{1,j-1}\Z\Crun{j,k-1}\Z\Crun{k,l-1}\Z\Crun{l,J}\\
\RBTaylorTerm{3}{1,2}  &=-\frac{1}{2}\sum_{j<k}(i\delta_j)\delta_k^2\Crun{1,j-1}\Z\Crun{j,k-1}\Z^2\Crun{k,J}\\
\RBTaylorTerm{3}{3}  &=-\frac{i}{6}\sum_{j=1}^J\delta_j^3\Crun{1,j-1}\Z^3\Crun{j,J}\\
\RBTaylorTerm{4}{1,1,1,1} & =\sum_{j<k<l<m}(i\delta_j)(i\delta_k)(i\delta_l)(i\delta_m)\Crun{1,j-1}\Z\Crun{j,k-1}\Z\Crun{k,l-1}\Z\Crun{l,m-1}\Z\Crun{m,J}\\
\RBTaylorTerm{4}{1,1,2}  &=-\frac{1}{2}
\sum_{j<k<l}(i\delta_j)(i\delta_k)\delta_l^2\Crun{1,j-1}\Z\Crun{j,k-1}\Z\Crun{k,l-1}\Z^2\Crun{l,J}\\
&
-\frac{1}{2}\sum_{j<k<l}(i\delta_j)\delta_k^2(i\delta_l)\Crun{1,j-1}\Z\Crun{j,k-1}\Z^2\Crun{k,l-1}\Z\Crun{l,J}\\
&
-\frac{1}{2}\sum_{j<k<l}\delta_j^2(i\delta_k)(i\delta_l)\Crun{1,j-1}\Z^2\Crun{j,k-1}\Z\Crun{k,l-1}\Z\Crun{l,J}
\\
\RBTaylorTerm{4}{2,2}  &=\frac{1}{4}\sum_{j<k}\delta_j^2\delta_k^2\Crun{1,j-1}\Z^2\Crun{j,k-1}\Z^2\Crun{k,J}\\
\RBTaylorTerm{4}{1,3}  &=-\frac{i}{6}\sum_{j<k}(i\delta_j)\delta_k^3\Crun{1,j-1}\Z\Crun{j,k-1}\Z^3\Crun{k,J}\\
\RBTaylorTerm{4}{4}  &=\frac{1}{24}\sum_{j=1}^J\delta_j^4\Crun{1,j-1}\Z^4\Crun{j,J}
\end{align}
and we have defined the Clifford \emph{subsequence} operators 
\begin{align}
\Crun{jk}\equiv\Clifford_{\CliffIndex{j}}...\Clifford_{\CliffIndex{k}},
\hspace{0.75cm}
1\le j\le k\le J.
\end{align} 
To evaluate Eq. \ref{Eq:TraceFidelityDefinition} we must obtain an expression for $\frac{1}{2}\text{Tr}(\RBseqNoisy_{\CliffVec,\NoiseVec})$. By the linearity of the trace, we must therefore obtain expressions for the quantities $\HTRBTaylorTerm{n}{k_1,k_2,...,k_m}\equiv\frac{1}{2}\Trace{\RBTaylorTerm{n}{k_1,k_2,...,k_m}}$ for each of the terms above. This requires some manipulation. To begin we highlight the following useful properties
\begin{align}\label{AppendixEq:ZkProperty}
&\Z^k = \begin{cases}
\Z,\hspace{0.75cm}&k\hspace{0.5cm}\text{odd}\\
\Identity,\hspace{0.75cm}&k\hspace{0.5cm}\text{even}
\end{cases}\\
&\Crun{1,J} \equiv \RBseqClean_{\CliffVec} \equiv \Identity.
\end{align}
In fact since $\Crun{1,J} = \Identity$, any cyclic permutation of subsequences $\Crun{jk}$ also gives the identity. For instance
\begin{align}
\Crun{1,J}=
\Crun{1,j-1}\Crun{j,k-1}\Crun{k,J}
=
\Crun{k,J}\Crun{1,j-1}\Crun{j,k-1}
\bla
=
\Crun{j,k-1}\Crun{k,J}\Crun{1,j-1}
\bla
=
\Identity
\end{align}
and so on for any number of subsequences. Using the cyclic property of the trace, we therefore have
\begin{align}
\frac{1}{2}\Trace{
\Crun{1,j-1}\Z^{k}\Crun{j,J}}=
\frac{1}{2}\Trace{
\Crun{j,J}\Crun{1,j-1}\Z^{k}}=
\frac{1}{2}\Trace{\Z^{k}}= \begin{cases}
0,\hspace{0.75cm}&k\hspace{0.5cm}\text{odd}\\
1,\hspace{0.75cm}&k\hspace{0.5cm}\text{even}
\end{cases}
\end{align}
The last equality uses Eq. \ref{AppendixEq:ZkProperty}, the property of Pauli matrices that $\text{Tr}(\Z) = 0$, and that the corresponding $(2\times2)$ identity has trace $\text{Tr}(\Identity) = 2$. Now define the cumulative operators $\Kstring{m}$ giving the product from the first through to the $m$th Clifford operator in the sequence
\begin{align}
\Kstring{m}\equiv\Crun{1,m} =
\Clifford_{\CliffIndex{1}}...\Clifford_{\CliffIndex{m}},
\hspace{1cm}
\Kstring{0}\equiv\Identity\equiv\Kstring{J}
\end{align}
where each $\Kstring{m}$ is some element of the Clifford group due to the closure property of groups under group composition. In this case any subsequence $\Crun{jk}$ ``factorizes'' into the products
\begin{align}
\Kstring{j-1}^\dagger
\Kstring{k} = 
\Clifford_{\CliffIndex{j-1}}^\dagger...\Clifford_{\CliffIndex{1}}^\dagger
\Clifford_{\CliffIndex{1}}...\Clifford_{\CliffIndex{j-1}}\Clifford_{\CliffIndex{j}}...\Clifford_{\CliffIndex{k}}
= \Crun{jk}
\end{align}
allowing us to rewrite
\begin{align*}
\Crun{1,j-1}
\Z
\Crun{j,k-1}
\Z
\Crun{k,J}
&
=
\PTwirl{j}
\PTwirl{k}
\\
\Crun{1,j-1}
\Z
\Crun{j,k-1}
\Z
\Crun{k,l-1}
\Z
\Crun{l,J}
&=
\PTwirl{j}
\PTwirl{k}
\PTwirl{l}
\\
\Crun{1,j-1}
\Z
\Crun{j,k-1}
\Z
\Crun{k,l-1}
\Z
\Crun{l,m-1}
\Z
\Crun{m,J}
&=
\PTwirl{j}
\PTwirl{k}
\PTwirl{l}
\PTwirl{m}\\
&\hspace{0.1cm}...
\end{align*}
where we define the $\Z$-conjugating operators 
\begin{align}
\label{AppendixEq:DefinitionPTwirl}
\PTwirl{m}\equiv\Kstring{m-1}\Z\Kstring{m-1}^\dagger\in\{\pm\X,\pm\Y,\pm\Z\},
\hspace{1.5cm}
0\le m \le J
\end{align}
and where $\PTwirl{m}$ is always a signed Pauli matrix due to the property that the Clifford group is the normalizer of the Pauli group. Thus we find 
\begin{align}
\HTRBTaylorTerm{0}{} &=1 \\
\HTRBTaylorTerm{1}{1} & =0\\
\HTRBTaylorTerm{2}{1,1}  &=-\frac{1}{2}\sum_{j<k}\delta_j\delta_k\Trace{\PTwirl{j}\PTwirl{k}}\\
\HTRBTaylorTerm{2}{2}  &=-\frac{1}{2}\sum_{j=1}^J\delta_j^2\\
\HTRBTaylorTerm{3}{1,1,1} & =-\frac{i}{2}\sum_{j<k<l}\delta_j\delta_k\delta_l\Trace{\PTwirl{j}\PTwirl{k}\PTwirl{l}}\\
\HTRBTaylorTerm{3}{1,2}  &=0\\
\HTRBTaylorTerm{3}{3}  &=0\\
\HTRBTaylorTerm{4}{1,1,1,1} & =\frac{1}{2}\sum_{j<k<l<m}\delta_j\delta_k\delta_l\delta_m\Trace{\PTwirl{j}\PTwirl{k}\PTwirl{l}\PTwirl{m}}\\
\HTRBTaylorTerm{4}{1,1,2}  &=\frac{1}{4}
\sum_{j<k<l}
\left[
\delta_j\delta_k\delta_l^2\Trace{\PTwirl{j}\PTwirl{k}}
+
\delta_j\delta_k^2\delta_l\Trace{\PTwirl{j}\PTwirl{l}}
+
\delta_j^2\delta_k\delta_l\Trace{\PTwirl{k}\PTwirl{l}}
\right]
\\
\HTRBTaylorTerm{4}{2,2}  &=\frac{1}{4}\sum_{j<k}\delta_j^2\delta_k^2\\
\HTRBTaylorTerm{4}{1,3}  &=\frac{1}{12}\sum_{j<k}\delta_j\delta_k^3\Trace{\PTwirl{j}\PTwirl{k}}\\
\HTRBTaylorTerm{4}{4}  &=\frac{1}{24}\sum_{j=1}^J\delta_j^4.
\end{align}
The nonzero terms may be recast into more convenient expressions by moving to vector notation. We expand the operators $\PTwirl{m}$ in the basis of Pauli operators,
\begin{align}
\PTwirl{m} = x_m\X+y_m\Y+z_m\Z,
\hspace{1cm}
x_m,y_m,z_m\in\{0,\pm1\},
\hspace{1cm}
\abs{x_m}^2+\abs{y_m}^2+\abs{z_m}^2=1.
\end{align} 
That is, only one nonzero coefficient $x_m,y_,z_m$, equivalent to expressing the fact that they are sampled from the set $\{\pm\X,\pm\Y,\pm\Z\}$. The associated unit vector
\begin{align}
\rhat_{m} \equiv (x_m,\hspace{0.05cm}y_m,\hspace{0.05cm}z_m),\hspace{1cm}\|\rhat_m\| = 1
\end{align}
therefore points uniformly at random along one of the principle Cartesian axes $\{\pm\xhat,\pm\yhat,\pm\zhat\}$, capturing the ``direction'' of the operator $\PTwirl{m}$ in ``Pauli space''.  With these definitions we may derive
\begin{align}
\frac{1}{2}\Trace{\PTwirl{j}\PTwirl{k}} &= \rhat_{j}\cdot\rhat_{k}\\
\frac{1}{2}\Trace{\PTwirl{j}\PTwirl{k}\PTwirl{l}} &= i\left(\rhat_{j}\times\rhat_{k}\right)\cdot\rhat_{l}\\
\frac{1}{2}\Trace{\PTwirl{j}\PTwirl{k}\PTwirl{l}\PTwirl{m}} &= 
\left(\rhat_{j}\cdot\rhat_{k}\right)\left(\rhat_{l}\cdot\rhat_{m}\right)
+\left(\rhat_{j}\cdot\rhat_{m}\right)\left(\rhat_{k}\cdot\rhat_{l}\right)
-\left(\rhat_{j}\cdot\rhat_{l}\right)\left(\rhat_{k}\cdot\rhat_{m}\right)
\end{align}
following directly from the trace and the cyclic composition properties of the Pauli matrices $\Pauli{i}\Pauli{j}=i\LeviCevita{ijk}\Pauli{k}+\delta_{ij}\Identity$,
where $\LeviCevita{ijk}$ is the Levi-Civita symbol, $\delta_{ij}$ is the Kronecker delta and Einstein summation notation used. 
Consequently we obtain  
\begin{align}
\HTRBTaylorTerm{2}{1,1}  &=-\sum_{j<k}\delta_j\delta_k \rhat_{j}\cdot\rhat_{k}\\
\HTRBTaylorTerm{2}{2}  &=-\frac{1}{2}\sum_{j=1}^J\delta_j^2\\
\HTRBTaylorTerm{3}{1,1,1} & =\sum_{j<k<l}\delta_j\delta_k\delta_l\left(\rhat_{j}\times\rhat_{k}\right)\cdot\rhat_{l}\\
\HTRBTaylorTerm{4}{1,1,1,1} & =\frac{1}{2}\sum_{j<k<l<m}\delta_j\delta_k\delta_l\delta_m
\left\{\left(\rhat_{j}\cdot\rhat_{k}\right)\left(\rhat_{l}\cdot\rhat_{m}\right)
+\left(\rhat_{j}\cdot\rhat_{m}\right)\left(\rhat_{k}\cdot\rhat_{l}\right)
-\left(\rhat_{j}\cdot\rhat_{l}\right)\left(\rhat_{k}\cdot\rhat_{m}\right)
\right\}
\\
\HTRBTaylorTerm{4}{1,1,2}  &=
\frac{1}{2}
\sum_{j<k<l}
\left[
\delta_j\delta_k\delta_l^2\rhat_{j}\cdot\rhat_{k}
+
\delta_j\delta_k^2\delta_l\rhat_{j}\cdot\rhat_{l}
+
\delta_j^2\delta_k\delta_l\rhat_{k}\cdot\rhat_{l}
\right]
\\
\HTRBTaylorTerm{4}{1,3}  &=\frac{1}{6}\sum_{j<k}\delta_j\delta_k^3\rhat_{j}\cdot\rhat_{k}\\
\HTRBTaylorTerm{4}{2,2}  &=\frac{1}{4}\sum_{j<k}\delta_j^2\delta_k^2\\
\HTRBTaylorTerm{4}{4}  &=\frac{1}{24}\sum_{j=1}^J\delta_j^4.
\end{align}
We expect the major contribution to the distribution of $\FidNoiseAv$ to reside in the term $\HTRBTaylorTerm{2}{1,1}$ since, in the limit $\sigma\ll1$, higher order terms  $\HTRBTaylorTerm{n>2}{}$ will be orders of magnitude smaller. Anticipating this, we recast $\HTRBTaylorTerm{2}{1,1}$ in terms of an unrestricted sum to facilitate more straightforward analysis. Observing the quantity $\delta_j\delta_k \rhat_{j}\cdot\rhat_{k}$ is invariant under exchange of indexes we thus obtain 
\begin{align}
\HTRBTaylorTerm{2}{1,1}& =-\frac{1}{2}\left(
\sum_{j=1}^J\sum_{k=1}^J\delta_j\delta_k \rhat_{j}\cdot\rhat_{k}-\sum_{j=1}^J\delta_j^2\rhat_{j}\cdot\rhat_{j}
\right)\\
&=-\frac{1}{2}\sum_{j,k}^J\delta_j\delta_k \rhat_{j}\cdot\rhat_{k}+\frac{1}{2}\sum_{j=1}^J\delta_j^2\\
&=-\frac{1}{2}\left(\sum_{j=1}^J\delta_j\rhat_{j}\right)\cdot\left(\sum_{k=1}^J\delta_k\rhat_{k}\right)-\HTRBTaylorTerm{2}{2}\\
&=-\frac{1}{2}\norm{\RvecA}^2-\HTRBTaylorTerm{2}{2}
\end{align}
where we define the resultant vector
\begin{align}
\RvecA\equiv\sum_{j=1}^{J}
\delta_{j}\rhat_{j}.
\end{align}
Hence 
\begin{align}
\label{AppendixEq:HTRBseqNoisyExpansion}
\frac{1}{2}\Trace{\RBseqNoisy_{\CliffVec,\NoiseVec}}&\approx 
1
-\frac{1}{2}\norm{\RvecA}^2
+\HTRBTaylorTerm{3}{1,1,1} 
+\HTRBTaylorTerm{4}{1,1,1,1} 
+\HTRBTaylorTerm{4}{1,1,2} 
+\HTRBTaylorTerm{4}{1,3} 
+\HTRBTaylorTerm{4}{2,2} 
+\HTRBTaylorTerm{4}{4}. 
\end{align}
Since the above terms are all real, the fidelity is then obtained taking the square of Eq. \ref{AppendixEq:HTRBseqNoisyExpansion} and truncating any cross terms higher than $\Order{\sigma^4}$. Thus we obtain
\begin{align}
\label{AppendixEq:FidelityExpansion1}
\mathcal{F}\left(\CliffVec,\NoiseVec\right)&\approx 
1
-\norm{\RvecA}^2
+\frac{1}{4}\left(\norm{\RvecA}^2\right)^2
+2\HTRBTaylorTerm{3}{1,1,1} 
+2\HTRBTaylorTerm{4}{1,1,1,1} 
+2\HTRBTaylorTerm{4}{1,1,2} 
+2\HTRBTaylorTerm{4}{1,3} 
+2\HTRBTaylorTerm{4}{2,2} 
+2\HTRBTaylorTerm{4}{4}. 
\end{align}
Averaging Eq. \ref{AppendixEq:FidelityExpansion1} over an ensemble of noise realizations $\NoiseVec$ then yields the approximate expression for $\FidNoiseAv$. However we make simplifying observation that only terms raised to even powers, or those summed over terms raised to even powers,  survive the ensemble average. Moreover, since the vectors  $\rhat_m$ are uniformly distributed over the set $\{\pm\xhat,\pm\yhat,\pm\zhat\}$, summing over compositions 
\begin{align}
&\left(\rhat_{j}\cdot\rhat_{k}\right)&&j< k<J\\
&\left(\rhat_{j}\times\rhat_{k}\right)\cdot\rhat_{l}&&j< k< l<J\\
&\left(\rhat_{j}\cdot\rhat_{k}\right)\left(\rhat_{l}\cdot\rhat_{m}\right)&&j< k< l< m<J
\end{align}
will on average resolve to the zero vector. Hence, even in the presence of correlated noise random variables, it is appropriate to set 
\begin{align}
\left\langle\HTRBTaylorTerm{3}{1,1,1} \right\rangle&\rightarrow0\\
\left\langle\HTRBTaylorTerm{4}{1,1,1,1} \right\rangle&\rightarrow0\\
\left\langle\HTRBTaylorTerm{4}{1,1,2} \right\rangle&\rightarrow0\\
\left\langle\HTRBTaylorTerm{4}{1,3} \right\rangle&\rightarrow0
\end{align}
In fact these quantities are random variables sharply (and symmetrically) peaked around 0. However since they are $\Order{\sigma^4}$ terms, the spread of their distributions is very small relative to the spread of $\langle\norm{\RvecA}^2\rangle$, which scales as $\Order{\sigma^2}$. The noise-averaged fidelity to $\Order{\sigma^4}$ therefore reduces to
\begin{align}
\label{AppendixEq:NAFExpansion1}
\FidNoiseAv&\approx 
1
-\left\langle\norm{\RvecA}^2\right\rangle
+O^{(4)}
\end{align}
where the final term in Eq. \ref{AppendixEq:NAFExpansion1} consists only of $\Order{\sigma^4}$ terms 
\begin{align}\label{AppendixEq:4thOrderResidualTerm}
O^{(4)} &\equiv \frac{1}{4}\left\langle{\norm{\RvecA}^4}\right\rangle
+2\left\langle\HTRBTaylorTerm{4}{2,2}\right\rangle
+2\left\langle\HTRBTaylorTerm{4}{4} \right\rangle.
\end{align}

The remaining task is to study how correlations between the noise variables affect the distributions of the terms in Eq. \ref{AppendixEq:NAFExpansion1}, and hence the distribution of $\FidNoiseAv$. In particular how these distributions change depending on whether the noise falls into Markovian or quasi-static regimes. As stated in the main text, our key insight is the interpretation of the vector quantity $\RvecA$ in terms of a 3D random walk generated by adding $J$ randomly-oriented steps with step lengths specified by $\NoiseVec$. The distribution of $\FidNoiseAv$ then maps onto the distance square of this 3D random walk. 

\bla
\newpage
\section{Universal Error Model}\label{Appendix:UniversalErrors}
\bla

\noindent
Here we sketch a generalization of our analytic framework, showing its applicability to \emph{universal errors}, beyond the (perhaps most important) case of dephasing specifically treated above.  As in the main text, errors are implemented by interleaving $\RBseqClean_{\CliffVec}$ with a sequence of stochastic unitary rotations, yielding the noise-affected operation
\begin{align}
\label{AppendixEq:RBNoisySequenceFullExpression}
\RBseqNoisy_{\CliffVec,\NoiseVec}&\equiv U_1\bla\Clifford_{\CliffIndex{1}}  
 U_2\bla\Clifford_{\CliffIndex{2}}  
\bla...
l U_J\bla\Clifford_{\CliffIndex{J}}.
\end{align}
However we now let the unitaries take the general form 
\begin{align}
U_j
\equiv \exp\left[-i\UniErrorVec_j\cdot\boldsymbol{\sigma}\right]
\equiv 
\exp\left[-i
\left(\UniError{j}{z}\Pauli{x}
+\UniError{j}{y}\Pauli{y}
+\UniError{j}{z}\Pauli{z}
\right)
\right]
\equiv 
\exp\left[-i
\left(\UniError{j}{z}\X
+\UniError{j}{y}\Y
+\UniError{j}{z}\Z
\right)
\right]
\end{align}
where $\boldsymbol{\sigma} \equiv (\Pauli{x},\Pauli{y},\Pauli{z}) \equiv(\X,\Y,\Z)$ denotes a vector of Pauli matrices and the vector $\UniErrorVec_j = (\UniError{j}{x},\UniError{j}{y},\UniError{j}{z})$ contains the error rotations induced about each Cartesian axis. That is, the unitary $U_j$ causes the net rotation through an angle $\norm{\UniErrorVec_j}$ about the axis $\hat{\boldsymbol{\delta}}_j \equiv \UniErrorVec_j/\norm{\UniErrorVec_j}$ on the Bloch sphere. In this case the error process indicates a list of 3-component \emph{vectors} 
\begin{align}
\NoiseVec = (
\UniErrorVec_1, \UniErrorVec_2,...,\UniErrorVec_J
).
\end{align}
Assuming the weak-noise limit, quantified by the perturbative condition $J\E{\norm{\UniErrorVec}^2}<1$, we make the Taylor approximation
\begin{align}
U_j\approx\Identity + i 
\left(\UniError{j}{z}\Pauli{x}
+\UniError{j}{y}\Pauli{y}
+\UniError{j}{z}\Pauli{z}
\right)
+...
\end{align}
In the main text we presented a complete treatment up to fourth order to account for both the leading- and higher-order contributions to infidelity, and assuming a dephasing-only environment.  Here we demonstrate rigorously how leading-order contributions to the fidelity take the same form when the noise model is universal, as described above.  Once again, the leading order contribution derives from the term like Eq. \ref{Eq:DominantXiTerm} but with a slight amendment: 
\begin{align}\label{AppendixEq:DominantXiTerm}
\RBTaylorTerm{2}{1,1}  &=\sum_{\alpha,\beta}\RBTaylorTermUni{2}{1,1}(\mu,\nu) 
\end{align}
where
\begin{align}
\RBTaylorTermUni{2}{1,1}(\mu,\nu)  &=-\sum_{j<k}\UniError{j}{\mu}\UniError{k}{\nu}\Crun{1,j-1}\Pauli{\mu}\Crun{j,k-1}\Pauli{\nu}\Crun{k,J}.
\end{align}
Writing $ \Crun{jk} = \Kstring{j-1}^\dagger \Kstring{k}$, and using $\Kstring{0}\equiv\Kstring{J}\equiv\Identity$, we therefore obtain 
\begin{align}\label{AppendixEq:DominantXiOperatorsInTermsOfPtwirls} 
\Crun{1,j-1}\Pauli{\mu}\Crun{j,k-1}\Pauli{\nu}\Crun{k,J}& =  \Kstring{0}^\dagger \Kstring{j-1}\Pauli{\mu} \Kstring{j-1}^\dagger \Kstring{k-1}\Pauli{\nu} \Kstring{k-1}^\dagger \Kstring{J}\\
& =\Kstring{j-1}\Pauli{\mu} \Kstring{j-1}^\dagger \Kstring{k-1}\Pauli{\nu} \Kstring{k-1}\\
& =\PTwirl{j}^{(\mu)}\PTwirl{k}^{(\nu)}
\end{align}
where the operators in the last equality generalize the definition in Eq.\ref{Eq:DefinitionPTwirl}, namely 
\begin{align}
\label{AppendixEq:DefinitionPTwirlUniversal}
\PTwirl{m}^{(\mu)}\equiv\Kstring{m-1}\Pauli{\mu}\Kstring{m-1}^\dagger\in\{\pm\X,\pm\Y,\pm\Z\}.
\end{align}
Once again, $0\le m \le J$, and the $\PTwirl{m}^{(\mu)}$ are always signed Pauli operators due to the property that the Clifford group is the normalizer of the Pauli group. We may therefore expand the operators in the basis of Pauli operators by writing
\begin{align}
\PTwirl{m}^{(\mu)} = x_m^{(\mu)}\X+y_m^{(\mu)}\Y+z_m^{(\mu)}\Z
\end{align}
where $x_m^{(\mu)},y_m^{(\mu)},z_m^{(\mu)}\in\{0,\pm1\}$ with only one nonzero coefficient. Since the Clifford sequences (and hence subsequences) are uniformly random, the operators $\PTwirl{m}^{(\mu)}$ are uniformly random, independent of the choice of $\mu$. The associated unit vector
\begin{align}
\rhat_{m}^{(\mu)} \equiv (x_m^{(\mu)},\hspace{0.05cm}y_m^{(\mu)},\hspace{0.05cm}z_m^{(\mu)}),\hspace{1cm}\|\rhat_m^{(\mu)}\| = 1
\end{align}
therefore still points uniformly at random along one of the principle Cartesian axes $\{\pm\xhat,\pm\yhat,\pm\zhat\}$. Taking the trace over Eq. \ref{AppendixEq:DominantXiOperatorsInTermsOfPtwirls}, using the cyclic composition properties of the Pauli matrices, and moving to vector notation we therefore obtain
\begin{align}
\frac{1}{2}\Trace{\PTwirl{j}^{(\mu)}\PTwirl{k}^{(\nu)}} &= \rhat_{j}^{(\mu)}\cdot\rhat_{k}^{(\nu)}.
\end{align}
Consequently 
\begin{align}
\HTRBTaylorTermUni{2}{1,1}(\mu,\nu) &\equiv\frac{1}{2}\Trace{\RBTaylorTermUni{2}{1,1}(\mu,\nu)  } = -\sum_{j<k}\UniError{j}{\mu}\UniError{k}{\nu}\rhat_{j}^{(\mu)}\cdot\rhat_{k}^{(\nu)}
\end{align}
where leading order contribution derives from the sum of all such terms $\HTRBTaylorTermUni{2}{1,1}(\mu,\nu)$
\begin{align}
\HTRBTaylorTerm{2}{1,1}  = \sum_{\mu,\nu = 1}^3\HTRBTaylorTermUni{2}{1,1}(\mu,\nu).
\end{align}
Taking an ensemble average over the noise realizations, we obtain 
\begin{align}
\left\langle\HTRBTaylorTerm{2}{1,1} \right\rangle &= \sum_{\mu,\nu = 1}^3\left\langle\HTRBTaylorTermUni{2}{1,1}(\mu,\nu) \right\rangle\\
&= -\sum_{\mu,\nu = 1}^3 
\sum_{j<k}
\left\langle\UniError{j}{\mu}\UniError{k}{\nu}\right\rangle
\rhat_{j}^{(\mu)}\cdot\rhat_{k}^{(\nu)}.
\end{align}
We now make a simplification in which we assume the errors in separate quadratures arises from distinct physical mechanisms (\emph{e.g.} dephasing and depolarization noise arise from different mechanisms). In this case the correlation between separate error quadratures is zero, and 
\begin{align}
\left\langle\UniError{j}{\mu}\UniError{k}{\nu}\right\rangle = \tilde{\delta}_{\mu,\nu}
\end{align}
where $\tilde{\delta}_{\mu,\nu}$ is the Kronecker delta. With this assumption, we obtain 
\begin{align}
\left\langle\HTRBTaylorTerm{2}{1,1} \right\rangle &= -\sum_{\mu = 1}^3 
\sum_{j<k}
\left\langle\UniError{j}{\mu}\UniError{k}{\mu}\right\rangle
\rhat_{j}^{(\mu)}\cdot\rhat_{k}^{(\mu)}.
\end{align}
Now the quantity $\UniError{j}{\mu}\UniError{k}{\mu}\rhat_{j}^{(\mu)}\cdot\rhat_{k}^{(\mu)}$ is invariant under exchange of lower indices, and we may recast the ordered sum as an unrestricted sum
\begin{align}
\left\langle\HTRBTaylorTerm{2}{1,1} \right\rangle &= -\frac{1}{2}\sum_{\mu = 1}^3 
\left[
\sum_{j,k}
\left\langle\UniError{j}{\mu}\UniError{k}{\mu}\right\rangle
\rhat_{j}^{(\mu)}\cdot\rhat_{k}^{(\mu)} 
-
\sum_{k=1}^{J}
\left\langle\left({\UniError{j}{\mu}}\right)^2\right\rangle
\right]
\end{align}
Defining
\begin{align}
\RvecA^{(\mu)}\equiv\sum_{i =1}
\UniError{j}{\mu}
\rhat_{i}^{(\mu)}
\end{align}
we obtain \begin{align}
\left\langle\HTRBTaylorTerm{2}{1,1} \right\rangle &= -\frac{1}{2}\sum_{\mu = 1}^3 
\left\langle
\norm{\RvecA^{(\mu)}}^2
\right\rangle
-
\left\langle
\HTRBTaylorTerm{2}{1,1} 
\right\rangle
\end{align}
where the residual term 
\begin{align}
\HTRBTaylorTerm{2}{2} = -\frac{1}{2}\sum_{\mu=1}^3\sum_{k=1}^J 
\left(\UniError{j}{\mu}\right)^2
\end{align}
is independent of the Clifford sequence and cancels out in the full expansion of $\FidNoiseAv$, as in the case of the dephasing error model in the main text. The leading order contribution to the noise-averaged \emph{infidelity} is therefore given by 
\begin{align}\label{AppendixEq:LeadingOrderErrorUniversal}
1-\FidNoiseAv\approx
\sum_{\mu = 1}^3 
\left\langle
\norm{\RvecA^{(\mu)}}^2
\right\rangle
\end{align}
where each of the terms $\langle\norm{\RvecA^{(\mu)}}^2\rangle$ are \emph{independent} random variable inheriting the PDF of the random walk, and taking the same form presented in the main text. Consequently each of the $\langle\norm{\RvecA^{(\mu)}}^2\rangle$ are mutually independent, gamma-distributed random variables. In general, each of these random walks will be distinct, and the sum in Eq. \ref{AppendixEq:LeadingOrderErrorUniversal} is over non-identical gamma-distributed random variables. The total PDF may be obtained from successive applications of Eq. \ref{Eq:SumOfIndependentRVs} by direct integration. In the event that all noise quadratures follow the \emph{same gamma-distribution} $\Gamma(\alpha,\beta)$, the total PDF simply reduces to the rescaled gamma distribution $\Gamma(3\alpha,\beta)$ by an application of Eq. \ref{AppendixEq:SumOfIndependentGammas}.


\bla
\newpage
\section{PDF Derivation - Markovian Regime}\label{Appendix:MarkovianDerivation}
\bla


\noindent
In the Markovian regime, we assume all noise random variables are i.i.d. Hence $\RvecA$ corresponds to a $J$-length unbiased random walk with step lengths sampled from the normal distribution $\mathcal{N}\left(0,\hspace{0.05cm}\sigma^2\right)$. Since these step lengths are symmetrically distributed about zero, the distributions of the components of the walk vector $\delta_j\rhat_{j} = (\delta_jx_{j},\hspace{0.05cm}\delta_jy_{j},\hspace{0.05cm}\delta_jz_{j})$ are invariant with respect to the sign of the coefficients $\alpha_{j}$ in all Cartesian directions $\alpha \in\{x,y,z\}$. Ignoring the signs we therefore treat the the coefficients as binaries $\alpha_j\in\{0,1\}$, where the zero event simply reduces the number of steps taken in that direction. Let 
\begin{align}
n_{\alpha} \equiv \sum_{i=1}^J\abs{\alpha_{j}},
\hspace{1cm}
\alpha\in\{x,y,z\},
\hspace{1cm}
n_x+n_y+n_z = J
\end{align}
count the total number of nonzero components in each Cartesian direction over the sequence of walk vectors $\{\rhat_{1},\rhat_{2},...,\rhat_{J}\}$. Thus
\begin{align}
\RvecA
&=\left(
\delta^{x}_1+...+\delta^{x}_{n_x},
\hspace{0.2cm}
\delta^{y}_1+...+\delta^{y}_{n_y},
\hspace{0.2cm}
\delta^{z}_1+...+\delta^{z}_{n_z}
\right)
\end{align}
where the superscripts in $\delta^{\alpha}_j$ indicate summing only over the subset of $\delta_j$ for which the coefficients $\alpha_{j}$ are nonzero. Thus we have 
\begin{align}
\norm{\RvecA}^2 =\Delta_x^2+\Delta_y^2+\Delta_z^2,
\hspace{1cm}
\Delta_\alpha\equiv
\left(\delta_1^{\alpha}+\delta_2^{\alpha}+...+\delta_{n_\alpha}^{\alpha}\right),\hspace{1cm}\alpha\in\{x,y,z\}.
\end{align}
Since all $\delta_j\sim\mathcal{N}\left(0,\hspace{0.05cm}\sigma^2\right)$ are i.i.d. in the Markovian regime, so too are all the random variables in set $S_\alpha\equiv\left\{\delta_1^{\alpha}, \delta_2^{\alpha},...\delta_{n_\alpha}^{\alpha}\right\}$, $\alpha\in\{x,y,z\}$. The distribution of their sum is  therefore given by 
\begin{align}
\Delta_{\alpha}\sim\mathcal{N}\left(0,\hspace{0.05cm}n_{\alpha}\sigma^2\right).
\end{align}
Further, since each $\rhat_{j}$ projects onto only a single Cartesian direction - and consequently the sets $S_{x,y,z}$ are mutually disjoint - the random variables $\Delta_{x,y,z}$ are mutually independent. The distribution of the sum of squares of non-identical Gaussians involves a generalized  generally challenging to write down, requiring a generalized chi-square distribution. To preserve analytic tractability, we make the following simplification. Since the vectors $\rhat_{j}$ are uniformly-distributed there is a $\frac{1}{3}$ probability of being parallel to any given Cartesian axis. The probability of getting any particular combination $(n_x,n_y,n_z)$ is therefore given by the multinomial distribution 
\begin{align}
\mathcal{P}\left(n_x,n_y,n_z\right) = \frac{J!}{n_x!n_y!n_z!}\left(\frac{1}{3}\right)^{n_x}\left(\frac{1}{3}\right)^{n_y}\left(\frac{1}{3}\right)^{n_z}, 
\hspace{1cm}
n_x+n_y+n_z = J
\end{align}
which, for $J\gtrapprox5$, is sufficiently peaked around $n_{x,y,z} = J/3$ that we may regard these values as fixed without significant error. In this case $\Delta_{x,y,z}\sim\mathcal{N}\left(0,\hspace{0.05cm}J\sigma^2/3\right)$ reduce to i.i.d. random variables. The distribution of $\norm{\RvecA}^2$ consequently reduces to chi-square distribution with 3 degrees of freedom. It is more convenient, however, to express this in more general terms as a member of the two-parameter family of gamma distributions (see Eq. \ref{AppendixEq:DefinitionGammaDistribution}), of which the chi-square is a special case. Specifically, we obtain the gamma distribution 
\begin{align}\label{Eq:RSQGammaDistribution}
&\norm{\RvecA}^2\sim\Gamma\left(\alpha,\beta\right),
\hspace{1cm}
\alpha = \frac{3}{2},
\hspace{1cm}
\beta=\frac{2J\sigma^2}{3}
\end{align}
with shape parameter $\alpha$ and scale parameter $\beta$. 
The distribution of a finite noise-ensemble average over $\norm{\RvecA}^2$ is therefore specified by  
\begin{align}
\langle \norm{\RvecA}^2\rangle_n = \frac{1}{n}\sum_{j=1}^n\norm{\RvecA}_j^2,
\hspace{1cm}\norm{\RvecA}_j^2\sim\Gamma\left(\frac{3}{2},\frac{2J\sigma^2}{3}\right)
\end{align}
where now the $\norm{\RvecA}_j^2$ are i.i.d. random variables. But the sample mean over $n$ gamma-distributed random variables simply yields a rescaled gamma distribution with $\alpha\rightarrow n\alpha$ and $\beta\rightarrow\beta/n$ (see Eq. \ref{AppendixEq:SampeMeanGammaDistributedRVs}). Consequently 
\begin{align}
\langle \norm{\RvecA}^2\rangle_n \sim\Gamma\left(\frac{3n}{2},\frac{2J\sigma^2}{3n}\right)
\end{align}
with moments 
\begin{align}
\E{\langle \norm{\RvecA}^2\rangle_n }&=J\sigma^2\\
\label{Eq:VarianceOfNoiseAveragedRSQ}\Var{\langle \norm{\RvecA}^2\rangle_n }&= \frac{2}{3}J^2\sigma^4n^{-1}.
\end{align}
From Eq. \ref{Eq:RSQGammaDistribution} the PDF for ${\norm{\RvecA}^2}$ now has the known form 
\begin{align}
f_{\norm{\RvecA}^2}(x) & = \frac{1}{\Gamma(\alpha)\beta^\alpha}x^{\alpha-1}\exp\left[-\frac{x}{\beta}\right]
\end{align}
where $\alpha = 3/2$, \hspace{0.25cm}$\beta = 2J\sigma^2/3$, and $\Gamma(x)$is the gamma function. The PDF for ${\norm{\RvecA}^4}$ therefore given by the transformation (see Eq. \ref{AppendixEq:PDFIncreasingFunctionOfRV})
\begin{align}
f_{\norm{\RvecA}^4}(x) &= \frac{1}{2\sqrt{x}}f_{\norm{\RvecA}^2}(\sqrt{x})
= 
\frac{3\sqrt{\frac{3}{2\pi}}e^{-\frac{3\sqrt{x}}{2J\sigma^2}}}{2(J\sigma^2)^{3/2}x^{1/4}}
\end{align}
By direct computation, the first two moments are then given by 
\begin{align}
\E{\norm{\RvecA}^4}& = \frac{5}{3}J^2\sigma^4\\
\Var{\norm{\RvecA}^4}& = \frac{80}{9}J^4\sigma^8
\end{align}
It is relevant at this point to consider the relative weight of the terms $\langle{\norm{\RvecA}^4}\rangle_n$ and $\langle{\norm{\RvecA}^2}\rangle_n$ in the calculation of $\FidNoiseAv$.  An application of the central limit theorem yields the approximation  
\begin{align}
\sqrt{n}\left(\langle {\norm{\RvecA}^4}\rangle_n-\E{\langle {\norm{\RvecA}^4}\rangle_n}\right)\sim\mathcal{N}\left(0,\hspace{0.05cm}\Var{{\norm{\RvecA}^4}}\right)
\end{align}
That is, $\langle\norm{\RvecA}^4\rangle_n$ is approximately Gaussian-distributed with mean and variance 
\begin{align}
\E{\langle\norm{\RvecA}^4\rangle_n}& =\E{\norm{\RvecA}^4} = \frac{5}{3}J^2\sigma^4\\
\label{Eq:VarianceOfNoiseAveragedRSQSQ}\Var{\langle\norm{\RvecA}^4\rangle_n}& =\Var{\norm{\RvecA}^4}/n= \frac{80}{9}J^4\sigma^8n^{-1}
\end{align}
The relative significance of these terms may be captured by the condition   $\Var{\langle\norm{\RvecA}^4\rangle_n}\lesssim\epsilon\Var{\langle\norm{\RvecA}^2\rangle_n}$ where $\epsilon$ is some small fraction. From Eqs. \ref{Eq:VarianceOfNoiseAveragedRSQ} and \ref{Eq:VarianceOfNoiseAveragedRSQSQ} this condition is met provided
\begin{align}
J\lesssim\sqrt{\frac{3\epsilon}{40}}\sigma^{-2}
\end{align}
For instance, if $\sigma \sim 0.01$, $\epsilon \lesssim 0.1$ provided we restrict $J\lesssim 1000$. Now, since $\FidNoiseAv$ involves a linear combination of the terms $\langle{\norm{\RvecA}^4}\rangle_n$ and $\langle{\norm{\RvecA}^2}\rangle_n$, the PDF of $\FidNoiseAv$ is approximately given by a convolution over the individual PDFs of these terms. However assuming $\epsilon$ is sufficiently small - so that the distribution of $\langle{\norm{\RvecA}^4}\rangle_n$ is sufficiently narrow - the primary contribution of this convolution is to shift the distribution by an amount approximately given by the mean of $\langle{\norm{\RvecA}^4}\rangle_n$. Thus we set 
\begin{align}
\langle{\norm{\RvecA}^4}\rangle_n\rightarrow\E{\langle\norm{\RvecA}^4\rangle_n} = \frac{5}{3}J^2\sigma^4
\end{align}
The other $\Order{\sigma^4}$ terms reduce to 
 \begin{align}
\left\langle\HTRBTaylorTerm{4}{2,2}\right\rangle
&=
\frac{1}{4}
\left\langle
\sum_{j<k}\delta_j^2\delta_k^2
\right\rangle
=
\frac{1}{4}\sum_{j<k}
\left\langle\delta_j^2\right\rangle
\left\langle\delta_k^2\right\rangle
=
\frac{1}{4}
\frac{J(J-1)}{2}\sigma^4
\\
\left\langle\HTRBTaylorTerm{4}{4}\right\rangle  
&=
\frac{1}{24}
\left\langle
\sum_{j=1}^J\delta_j^4
\right\rangle
=
\frac{1}{24}\sum_{j=1}^J\left\langle\delta_j^4
\right\rangle
=
\frac{1}{24}J(3\sigma^4)
\end{align}
\bla
Substituting in these constants, Eqs. \ref{AppendixEq:NAFExpansion1} and \ref{AppendixEq:4thOrderResidualTerm}, the noise average fidelity for Markovian errors reduces to 
\begin{align}
\label{AppendixEq:NAFfinalMK}
\FidNoiseAv&\approx 
1
-\left\langle\norm{\RvecA}^2\right\rangle_n
+\frac{2}{3}J^2\sigma^4,
\hspace{1cm}
\langle \norm{\RvecA}^2\rangle_n \sim\Gamma\left(\frac{3n}{2},\frac{2J\sigma^2}{3n}\right)
\end{align}
Performing the appropriate linear transformations (Eq. \ref{Eq:Ch6PDFofLTofRV}) to incorporate the constant factors in Eq.\ref{AppendixEq:NAFfinalMK}, and using the definition of the PDF for a gamma-distribution (Eq. \ref{AppendixEq:DefinitionGammaDistribution}), the PDF for $\FidNoiseAv$ finally takes the form 
\begin{align}
f_{\FidNoiseAv}(F)
& = \frac{1}{\Gamma(\alpha)\beta^\alpha}\nu(F)^{\alpha-1}\exp\left[-\frac{\nu(F)}{\beta}\right]\\
&\nu(F)=1-F+ \frac{2}{3}J^2\sigma^4\\
&\alpha = 3n/2,\hspace{0.75cm}\beta = 2J\sigma^2/3n
\end{align}
where $\Gamma(x)$is the gamma function.

\bla
\newpage
\section{PDF Derivation - Quasi-Static (DC) Regime}\label{Appendix:DCDerivation}
\bla
%
In the DC regime we assume all noise random variables $\delta_j\equiv\delta$ are, in a given instance, identical (maximally correlated). However over separate instances $\delta$ is sampled from the normal distribution $\delta\sim\mathcal{N}\left(0,\hspace{0.05cm}\sigma^2\right)$. In a given run, the random walk vector $\RvecA=\sum_{j=1}^{J}
\delta\rhat_{j}
=\delta\sum_{j=1}^{J}
\rhat_{j}$ therefore reduces to a $J$-length unbiased random walk on a 3D lattice with \emph{fixed} step length $\delta$. 
In this case the noise random variables $\delta$ and Clifford-dependent random variables $\rhat_{j}$ factorize, allowing us to express 
\begin{align}
\RvecA=\delta\Vvec,
\hspace{1.5cm}
\Vvec\equiv\sum_{j=1}^J\rhat_{j}
\end{align}
where $\Vvec\in\mathbb{R}^3$ defines an unbiased random walk on a 3D lattice generated by adding $J$ \emph{unit-length} steps. Since we are interested in the norm square
$\norm{\RvecA}^2 = \delta^2\norm{\Vvec}^2$, however,
any sign dependence of $\delta$ vanishes. Performing a finite ensemble average over noise randomizations we therefore obtain 
\begin{align}
\langle\norm{\RvecA}^2\rangle_n 
= 
\langle\delta^2\rangle_n
\norm{\Vvec}^2,
\hspace{1.5cm} 
\langle\norm{\RvecA}^4\rangle_n
= 
\langle\delta^4\rangle_n
\norm{\Vvec}^4
\end{align}
Substituting these expressions into Eq. \ref{AppendixEq:NAFExpansion1} we therefore obtain 
$\FidNoiseAv\approx 
1
-\langle\delta^2\rangle_n\norm{\Vvec}^2
+
\frac{1}{4}\langle\delta^4\rangle_n\norm{\Vvec}^4
+
\Order{\sigma^4}$
where the term $\Order{\sigma^4}$ includes the quantities $\langle\HTRBTaylorTerm{4}{2,2}\rangle$ and $\langle\HTRBTaylorTerm{4}{2,2}\rangle$, which approximately reduce to constants. 
%
Since the term in $\norm{\Vvec}^2$ and $\norm{\Vvec}^4$ are now \emph{highly correlated}, however, we cannot exploit simplifying properties such as approximate independence between primary and higher-order terms in the expansion. One approach would involve making the approximation that $\langle\delta^2\rangle_n\rightarrow \sigma^2$ and $\langle\delta^4\rangle_n\rightarrow 3\sigma^4$ for sufficiently large $n$, and then 
completing the square in $\norm{\Vvec}^2$. In this case the PDF for $\FidNoiseAv$ could be obtained by successively performing linear, square and shifting transformations on the PDF of $\norm{\Vvec}^2$. However to a good approximation it turns out most of the physics is captured by the first term $\langle\delta^2\rangle_n\norm{\Vvec}^2$. Hence we make the approximation
\begin{align}\label{AppendixEq:NAFtruncationDC1}
\FidNoiseAv&\approx 
1
-\langle\delta^2\rangle_n\norm{\Vvec}^2
\end{align} 
As shown in the main text, proceeding with this truncating the expansion produces good agreement with direct simulation. 
The PDF for $\FidNoiseAv$ is therefore obtained by incorporating the distributions of $\norm{\Vvec}^2$ and $\langle\delta^2\rangle_n$.
The PDF of $\delta$ is given by the Gaussian $f_{\delta}(x) := \frac{1}{\sqrt{2\pi}\sigma}\exp\left(-\frac{x^2}{2\sigma^2}\right)$. Hence the PDF for $\delta^2$ is given by the transformation (see Eq. \ref{AppendixEq:PDFIncreasingFunctionOfRV})
\begin{align}
f_{\delta^2}(x) &= 2\left[\frac{1}{2 x^{1/2}}f_{\delta}\left(x^{-1/2}\right)\right]\\
&= \frac{1}{\sqrt{\pi}}\frac{1}{\sqrt{2\sigma^2}}x^{\frac{1}{2}-1}\exp\left(-\frac{x}{2\sigma^2}\right)\\
&= \frac{1}{\Gamma(\frac{1}{2})}\frac{1}{(2\sigma^2)^{1/2}}x^{\frac{1}{2}-1}\exp\left(-\frac{x}{2\sigma^2}\right)\\
&= \frac{1}{\Gamma(\alpha)\beta^\alpha}x^{\alpha-1}\exp\left(-\frac{x}{\beta}\right)
\end{align}
where $\alpha = 1/2$ and $\beta = 2\sigma^2$, and $\Gamma(x)$  is the gamma function. Comparing this with Eq. \ref{AppendixEq:DefinitionGammaDistribution}, this is the PDF of a Gamma distribution with shape parameter $\alpha$ and scale parameter $\beta$. The distribution of a finite noise-ensemble average over $\delta_2$ is therefore specified by  
\begin{align}
\langle\delta^2\rangle_n = \frac{1}{n}\sum_{j=1}^n\delta^2_j,
\hspace{1cm}\delta^2_j\sim\Gamma\left(\frac{1}{2},\hspace{0.05cm}2\sigma^2\right)
\end{align}
where the $\delta_j^2$ are i.i.d. random variables. Now the sample mean over $n$ gamma-distributed random variables simply yields a rescaled gamma distribution with $\alpha\rightarrow n\alpha$ and $\beta\rightarrow\beta/n$ (see Eq. \ref{AppendixEq:SampeMeanGammaDistributedRVs}). Consequently 
\begin{align}
\langle\delta^2\rangle_n \sim\Gamma\left(\frac{n}{2},\frac{2\sigma^2}{n}\right)
\end{align}
with moments 
\begin{align}
\E{\langle \delta^2\rangle_n }&=\sigma^2\\
\hspace{1.5cm}
\Var{\langle \delta^2\rangle_n }&= \frac{2\sigma^4}{n}
\end{align}
As outlined above, $\norm{\Vvec}^2$ expresses the distance square of an unbiased random walk on a 3D lattice generated by adding $J$ \emph{unit-length} steps. Let $R$ be the random variable representing the distance from the origin in a symmetric (Bernoulli) 3D random walk after $J$ steps. Then the PDF for $R$ is known to be
\begin{align}
\label{AppendixEq:BernoulliRandomWalkDistancePDF}
f_{R}(r) = \left(\frac{3}{2\pi J}\right)^{3/2}4\pi r^2 e^{\frac{-3r^2}{2J}}
\end{align}
The distribution of the distance square $\norm{\Vvec}^2 = R^2$ is therefore given by the transformation (see Eq. \ref{AppendixEq:PDFIncreasingFunctionOfRV})
\begin{align}
f_{\norm{\Vvec}^2}(x) &= \frac{1}{2 x^{1/2}}f_{R}\left(x^{-1/2}\right)\\
&=\left(\frac{3}{2\pi J}\right)^{3/2}2\pi x^{1/2}\exp\left[\frac{-3x}{2J}\right]\\
&=\left(\frac{2J}{3}\right)^{-3/2}
\left(\frac{\sqrt{\pi}}{2}\right)^{-1}
x^{1/2}\exp\left[\frac{-x}{2J/3}\right]\\
&=\left(\frac{2J}{3}\right)^{-3/2}
\Gamma\left(\frac{3}{2}\right)^{-1}
x^{1/2}\exp\left[\frac{-x}{2J/3}\right]\\
&= \frac{1}{\Gamma(\alpha)\beta^\alpha}x^{\alpha-1}\exp\left(-\frac{x}{\beta}\right)
\end{align}
where $\alpha = 3/2$ and $\beta = 2J/3$, and $\Gamma(x)$  is the gamma function. But this is the PDF of a Gamma distribution with shape parameter $\alpha$ and scale parameter $\beta$. Consequently 
\begin{align}
\norm{\Vvec}^2\sim\Gamma\left(\frac{3}{2},\hspace{0.05cm}\frac{2J}{3}\right)
\end{align}
with moments 
\begin{align}
\E{\norm{\Vvec}^2}&=J\\
\hspace{1.5cm}
\Var{\norm{\Vvec}^2 }&= \frac{2J^2}{3}
\end{align}
Thus, to first order the PDF for $\FidNoiseAv$ is specified by the product of independent gamma-distributed random variables. This class of distributions is generally difficult to express. However we may obtain a closed form for the PDF of $\langle\delta^2\rangle_n\norm{\Vvec}^2$ by direct integration (see Eq. \ref{AppendixEq:ProductDistribution}), yielding
\begin{align}\label{AppendixEq:NAFDCPDFFiniteSampling}
f_{\FidNoiseAv} &= \frac{
\kappa^{\frac{n+3}{4}}
\left(\frac{\nu}{4}\right)^{\frac{n-1}{4}}
K_{\frac{1}{2}(n-3)}
\left(
\sqrt{\kappa\nu}
\right)
}
{\Gamma\left(\frac{J}{2}\right)}
\end{align}
where $\nu=1-F$, 
$\kappa=3n/J\sigma^2$
and
$K_n(z)$ gives the modified Bessel function of the second kind. However for fairly reasonable ensemble sizes $n\gtrsim 50$ it is sufficient to approximate $\langle\delta^2\rangle_n$ as the mean of its distribution, namely $\sigma^2$. In this case the fidelity distribution reduces simply to the (scaled and shifted) gamma distribution associated with  which $\norm{\Vvec}^2$, yielding
\begin{align}
\label{AppendixEq:NAFDCPDFInfiniteSampling}
f_{\FidNoiseAv}(F)
& = \frac{1}{\Gamma(\alpha)\beta^\alpha}\nu^{\alpha-1}\exp\left[-\frac{\nu}{\beta}\right]\\
\nu&=1-F\\
\alpha &= 3/2,\hspace{0.75cm}\beta = 2J\sigma^2/3
\end{align}

\bla
\newpage
\section{Probability distribution functions for partially correlated noise}\label{Sec:BlockCorrelated}
\bla

In the main text we derived gamma distributions for $\FidNoiseAv$ in the Markovian and DC limits, and assumed continuity of the distribution to interpolate between these extremal cases.   In this appendix we present formal justification of this approach. We treat a specific intermediate-correlation-length model where errors are \emph{block-correlated}, \emph{i.e.}\ within blocks of length $M$, the noise value is identical, and there is no correlation between blocks.  That is, each noise realization, $\NoiseVec$, can be partitioned into blocks of length $M$, within which the $\delta_i \sim\mathcal{N}(0,\sigma^2)$ are constant.  In block $k$, the corresponding random walk takes steps along the cartesian axes (in either direction), which we count as $m_k^{+x},m_k^{-x},m_k^{+y},m_k^{-y},m_k^{+z},m_k^{-z}$.  The step counts satisfy the constraint 

\begin{align}
m_k^{+x}+m_k^{-x}+m_k^{+y}+m_k^{-y}+m_k^{+z}+m_k^{-z}=M
\end{align}
and are multinomially distributed,
\begin{align}
{\boldsymbol m}_k\equiv(m_k^{+x},m_k^{-x},m_k^{+y},m_k^{-y},m_k^{+z},m_k^{-z})\sim\mathcal{M}(M,\{\frac{1}{6},\frac{1}{6},\frac{1}{6},\frac{1}{6},\frac{1}{6},\frac{1}{6}\}). 
\end{align}
The displacement vector associated with block $k$ is then 
\begin{align}
\RvecA_k &=\delta_k \big(m_k^{+x}-m_k^{-x},m_k^{+y}-m_k^{-y},m_k^{+z}-m_k^{-z}\big)\nonumber\\
&\equiv \delta_k\Vvec_k
\end{align}
where  $\Vvec_k$ is the displacement vector of a random walk associated with the $k$th block, involving $M$ \emph{unit-length} steps along the cartesian axes. The total displacement is then
\begin{align}
\RvecA =\sum_{k=1}^N \RvecA_k,\hspace{1cm}N=J/M
\end{align}
where $N$ is the number of $M$-length blocks in a sequence of total length $J$. In this picture the Markovian limit corresponds to the case  $M=1, N=J$ ($J$ blocks, each consisting of a single Clifford gate); the DC limit corresponds to the case  $M=J, N=1$ (1 block of length J). We therefore obtain 
\begin{align}
\norm{\RvecA}^2 
&=\sum_{i=1}^N\sum_{j=1}^N \RvecA_i\cdot\RvecA_j\\
&=\sum_{k=1}^N \norm{\RvecA_k}^2+\sum_{i\ne j}^N \RvecA_i\cdot\RvecA_k\\
&=\sum_{k=1}^N \delta_k^2 \norm{\Vvec_k}^2+\sum_{i\ne j}^N \delta_i\delta_j\Vvec_i\cdot\Vvec_j.
 \end{align}
Taking the ensemble average over noise realizations then yields
\begin{align}\label{AppendixEq:RsqAvBlock1}
\langle\norm{\RvecA}^2 \rangle
&=\sum_{k=1}^N \langle \delta_k^2\rangle \norm{\Vvec_k}^2+\sum_{i\ne j}^N \langle \delta_i\delta_j\rangle\Vvec_i\cdot\Vvec_j.
 \end{align}
Here the angle brackets $\langle\cdot\rangle$ refer to an ensemble average over noise realizations in the limit of an \emph{infinite} ensemble size ($n\rightarrow\infty$). From our assumption that the noise is perfectly block correlated (\emph{i.e.} zero correlation between errors from different blocks), and Gaussian-distributed within each block $\delta_i \sim\mathcal{N}(0,\sigma^2)$, we therefore obtain 
 \begin{align}
 \langle\delta_i\delta_j\rangle = \sigma^2\tilde{\delta}_{ij}
 \end{align}
 where $\tilde{\delta}_{ij}$ is the Kronecker delta. Thus, the second term in Eq. \ref{AppendixEq:RsqAvBlock1} vanishes, yielding
\begin{align}\label{AppendixEq:RsqAvBlock2}
\langle\norm{\RvecA}^2 \rangle
=\sum_{k=1}^N \sigma^2 \norm{\Vvec_k}^2
=\sigma ^2\sum _{k=1}^N\left[
{(m_k^{+x}-m_k^{-x})^2+(m_k^{+y}-m_k^{-y})^2+(m_k^{+z}-m_k^{-z}) ^2)}\right]
 \end{align}
 \noindent In general, $\langle\norm{\RvecA}^2\rangle$, depends on the details of the Clifford sequence $\CliffVec$ through its dependence  on the $\Vvec_k$, or equivalently, on the step counts $\bf{m}_k$.  Considering an ensemble of Clifford sequences, we can compute the statistics of $\langle\norm{\RvecA}^2\rangle$. The first two moments are straightforward to calculate exactly. We find  
 \begin{align}
 \E{((m_k^{+x}-m_k^{-x})^2+(m_k^{+y}-m_k^{-y})^2+(m_k^{+z}-m_k^{-z}) ^2}&=\sigma^2 M,\nonumber\\
 \Var{(m_k^{+x}-m_k^{-x})^2+(m_k^{+y}-m_k^{-y})^2+(m_k^{+z}-m_k^{-z}) ^2)}&=2\,\sigma^4 M(M-1)/3,\nonumber
 \end{align}
which, together with Eq. \ref{AppendixEq:RsqAvBlock2}, therefore yields  
 \begin{align}
\label{AppendixEq:ExpectationMcorrelated}\E{\langle\norm{\RvecA}^2\rangle}&=\sigma^2 J,\\
\label{AppendixEq:VarianceMcorrelated}\Var{\langle\norm{\RvecA}^2\rangle}&=2\,\sigma^4 J(M-1)/3.
\end{align}

In the Markovian limit ($M=1, N=J$), the expectation $\E{\langle\norm{\RvecA}^2\rangle}=\sigma^2 J$ is in fact independent of the details of the sequence, since $\bf{m}$ consists of a single unit entry, so that the summand in \eqn{AppendixEq:RsqAvBlock2} is identically 1.  This is perhaps not surprising: if different Clifford sequences had different noise-averaged fidelities for Markovian noise, then the optimal sequences would be good candidates for dynamical-decoupling sequences to suppress Markovian noise.  It is known that dynamical decoupling is not useful for correcting fast noise. The variance is  consistent with the $n\rightarrow\infty$ limit,  $\Var{\langle\norm{\RvecA}^2\rangle_\infty}=0$ in the main text (see Table \ref{Table:T1}). For DC noise  ($M=J, N=1$), we find the same Clifford-averaged expectation as above, however 
\begin{align}
\Var{\langle\norm{\RvecA}^2\rangle}=2\sigma^4 J(J-1)/3\approx 2\sigma^4 J^2/3
\end{align}
in agreement with earlier results in this limit up to a term $\Order{\sigma^4}$ (see Table \ref{Table:T1}). 

In a given experimental scenario, it may be  that there is a maximum correlation length $\Mmax$, beyond which gates are uncorrelated.  The above results then suggests that the clifford-averaged variance will vary quadratically with $J$ for $J\ll\Mmax$, and linearly with $J$ for $J\gg\Mmax$, with a transition around $J=\Mmax$.  This may be a useful heuristic for identifying correlation lengths in experiments.

For a given sequence length, $J$, the PDF of $\langle\norm{\RvecA}^2\rangle$ is in fact discrete.  However, for modest sizes of $J$ and $M$ we can approximate the PDF by a continuous distribution.  Given that the PDFs in both the DC and Markovian limits are both well-approximated by gamma distributions, we can use the calculated expectation and variance above to guess the PDF in the regime of intermediate correlation length.  Since $\E{\Gamma(\alpha,\beta)}=\alpha \beta$ and $\Var{\Gamma(\alpha,\beta)}=\alpha \beta^2$, and using Eqs. \ref{AppendixEq:ExpectationMcorrelated} and \ref{AppendixEq:VarianceMcorrelated}, we guess
\begin{align}
\langle\norm{\RvecA}^2\rangle&\sim\Gamma\Big(\frac{3J}{2(M-1)},\frac{2(M-1)\sigma^2}{3}\Big),\label{Rgood}\\
&\approx \Gamma\left(\frac{3J}{2M},\hspace{0.05cm}\frac{2M\sigma^2}{3}\right),\quad \textrm{for $M\gg1$} \label{Rapprox}
\end{align}
\eqn{Rapprox} interpolates between the two limiting cases. 

This guess can formalized by recalling $\langle\norm{\RvecA}^2 \rangle=\sum_{k=1}^N \sigma^2 \norm{\Vvec_k}^2$.  However, each of the $\Vvec_k$ represent the displacement of a \emph{independent} random walk taking $M$ steps of unit length along the cartesian axes, and are formally identical to the random walk vector in the DC limit. From Eq. \ref{Eq:DistributionOfVvec} the displacement square of this random walk, given $J$ steps, is given by the gamma distribution $ \norm{\Vvec}^2\sim\Gamma\left(\frac{3}{2},\hspace{0.05cm}\frac{2J}{3}\right)$. Consequently Eq. \ref{AppendixEq:RsqAvBlock2} describes the sum, scaled by $\sigma^2$, of $N$ \emph{independent and identically distributed} random variables, each following the distribution 
\begin{align}
\norm{\Vvec_k}^2\sim\Gamma\left(\frac{3}{2},\hspace{0.05cm}\frac{2M}{3}\right).
\end{align}
By an application of Eqs. \ref{AppendixEq:SumOfIndependentGammas} and \ref{AppendixEq:SampeMeanGammaDistributedRVs} we therefore recover 
\begin{align}
\langle\norm{\RvecA}^2 \rangle\sim\Gamma\left(\frac{3J}{2M},\hspace{0.05cm}\frac{2M\sigma^2}{3}\right)
\end{align}
which is asymptotically correct for large $M$. In showing this we have demonstrated analytically that intermediate correlation structures between the Markovian and DC limits are also described by the gamma distribution.

\bla
\newpage
\section{Fidelity Statistics for Generically-Correlated Processes}\label{Sec:STATSforGenericCorrelations} 
\bla

\noindent
Here we derive the expectation $\E{\langle\norm{\RvecA}^2\rangle}$ and variance $\Var{\langle\norm{\RvecA}^2\rangle}$ stated in Eq. \ref{Eq:StatsForACfunction} of the main text, for an error process $\NoiseVec$ with generic correlation structure specified by an autocorrelation function. Consider an arbitrary time-series $x(t)$ describing some (continuous) time-dependent, wide-sense stationary error process. Then $x(t)$ may be characterized by an autocorrelation function
\begin{align}
C_x(\tau) = \left\langle x(t)x(t+\tau)\right\rangle_t = \lim_{T\rightarrow\infty}\frac{1}{2T}\int_{-T}^{T}x(t)x(t+\tau)
\end{align}
where $\langle\rangle_t$ refers to an ensemble average over the time series and $\tau$ is the \emph{time difference} between measurements. Invoking the Wiener-Khintchine theorem, $C_x(\tau)$ and the power spectral-density (PSD) $S(\omega)$ form a Fourier-transform pair
\begin{align}
S(\omega) = \frac{1}{\sqrt{2\pi}}\int_{-\infty}^{\infty}C(\tau)e^{i\omega\tau}d\tau
\iff
 C(\tau) = \int_{-\infty}^{\infty}S(\omega)e^{-i\omega\tau}d\omega
\end{align}
We make use of these relations by discretizing the time-series, whereby 
\begin{align}
\RvecA=\sum_{j=1}^{J}
\delta_{j}\rhat_{j},\hspace{1cm}\delta_j=x(t_j),\hspace{1cm}t_j/\tau_g \in \{1,2,..,J\}
\end{align}
and $\tau_{g}$ is the time taken to perform a Clifford operation. The underlying error process is thereby discretely ``sampled'' by the Clifford sequence $\RBseqClean_{\CliffVec}$, and correlations between elements of $\NoiseVec$ separated by a time interval of ``$k$ gates'' are specified by the (discrete) autocorrelation function
\begin{align}
C_{\NoiseVec}(k) \equiv \left\langle \delta_j \delta_{j+k}\right\rangle_j
\end{align}
\subsubsection*{Expectation}
\noindent
We begin by expanding 
\begin{align}
\langle\RnormSQ\rangle = \langle\RvecA\cdot\RvecA\rangle
 = \sum_{i=1}^{J}\sum_{j=1}^{J}
\langle x(t_i)x(t_j)\rangle\rhat_{i}\cdot\rhat_{j}.
\end{align}
Here the ensemble average indicated by $\langle\cdot\rangle$ is over the noise random variables, and so does not affect the Clifford-dependent unit vectors $\rhat_i$. To obtain the full expectation we now take a separate expectation over the the random variables $\rhat_i$, which does not affect the noise random variables. Since the Clifford unit vectors $\rhat_{i}$ are uniformly and randomly distributed over the set $\{\pm\xhat,\pm\yhat,\pm\zhat\}$ the expectation over the inner product $\E{\rhat_{i}\cdot\rhat_{j}} = \tilde{\delta}_{ij}$ (Kronecker delta). Hence 
\begin{align}
\E{\langle\RnormSQ\rangle} 
&= \sum_{i=1}^{J}\sum_{j=1}^{J}
\langle x(t_i)x(t_j)\rangle\E{\rhat_{i}\cdot\rhat_{j}}\\
&= \sum_{i=1}^{J}\sum_{j=1}^{J}
\langle x(t_i)x(t_j)\rangle \tilde{\delta}_{ij}\\
&= \sum_{i=1}^{J}
\langle x(t_i)x(t_i)\rangle \\
&= \sum_{i=1}^{J}
\langle \delta_i\delta_i\rangle \\
&=  \sum_{i=1}^{J}C_{\NoiseVec}(0)\\
&=  JC_{\NoiseVec}(0)
\end{align}

\subsubsection*{Variance}

\noindent For the variance we need to find the expectation of $\langle\RnormSQ\rangle^2$. Using the above result, we momentarily define
\begin{align}
\mathcal{E} \equiv \E{\langle\RnormSQ\rangle} 
 = \sum_{k=1}^{J}
\langle \delta_k\delta_k\rangle
 = \sum_{k=1}^{J}
\langle \delta_k^2\rangle
\end{align}
Now we rewrite
\begin{align}
\langle\RnormSQ\rangle&=
\sum_{k=1}^{J}
\langle \delta_{k}^2\rangle
+
\sum_{i\ne j}^{J}
\langle \delta_{i}\delta_{j}\rangle
(\rhat_{i}\cdot\rhat_{j})
=
\mathcal{E}
+
\sum_{i\ne j}^{J}
\langle \delta_{i}\delta_{j}\rangle
(\rhat_{i}\cdot\rhat_{j})
\end{align}
Hence 
\begin{align}
(\langle\RnormSQ\rangle)^2&=
\left(
\mathcal{E}
+
\sum_{i\ne j}^{J}
\langle \delta_{i}\delta_{j}\rangle
(\rhat_{i}\cdot\rhat_{j})
\right)
\left(
\mathcal{E}
+
\sum_{i'\ne j'}^{J}
\langle \delta_{i'}\delta_{j'}\rangle
(\rhat_{i'}\cdot\rhat_{j'})
\right)\\
&=
\mathcal{E}^2
+
2\mathcal{E}^2\sum_{i\ne j}^{J}
\langle \delta_{i}\delta_{j}\rangle
(\rhat_{i}\cdot\rhat_{j})
+
\sum_{i\ne j}^{J}
\sum_{i'\ne j'}^{J}
\langle \delta_{i}\delta_{j}\rangle
\langle \delta_{i'}\delta_{j'}\rangle
(\rhat_{i}\cdot\rhat_{j})
(\rhat_{i'}\cdot\rhat_{j'})
\end{align}
As above, $\E{\rhat_{i}\cdot\rhat_{j}} = \tilde{\delta}_{ij}$, so taking the expectation over the the random variables $\rhat_i$ causes the term $2\mathcal{E}^2\sum_{i\ne j}^{J}
\langle \delta_{i}\delta_{j}\rangle
(\rhat_{i}\cdot\rhat_{j})$ vanishes, yielding
\begin{align}
\E{(\langle\RnormSQ\rangle)^2}=
\mathcal{E}^2
+
\sum_{i\ne j}^{J}
\sum_{i'\ne j'}^{J}
\langle \delta_{i}\delta_{j}\rangle
\langle \delta_{i'}\delta_{j'}\rangle
\E{
(\rhat_{i}\cdot\rhat_{j})
(\rhat_{i'}\cdot\rhat_{j'})
}
\end{align}
Hence the variance is given by 
\begin{align}
\Var{\langle\RnormSQ\rangle} 
&= \E{(\langle\RnormSQ\rangle)^2} - \E{(\langle\RnormSQ\rangle)}^2\\
&= \E{(\langle\RnormSQ\rangle)^2} -\mathcal{E}^2\\
&=\sum_{i\ne j}^{J}
\sum_{i'\ne j'}^{J}
\langle \delta_{i}\delta_{j}\rangle
\langle \delta_{i'}\delta_{j'}\rangle
\E{
(\rhat_{i}\cdot\rhat_{j})
(\rhat_{i'}\cdot\rhat_{j'})
}
\end{align}
Now for $i\ne j$ and $i' \ne j'$, the expectation over $(\rhat_{i}\cdot\rhat_{j})(\rhat_{i'}\cdot\rhat_{j'})$ is zero unless either 
\begin{align}
&i = i' \hspace{0.1cm}\&\hspace{0.1cm} j = j'\hspace{2cm}\implies\hspace{2cm}\E{(\rhat_{i}\cdot\rhat_{j})(\rhat_{i'}\cdot\rhat_{j'})}=1/3\\
&i = j' \hspace{0.1cm}\&\hspace{0.1cm} j = i'\hspace{2cm}\implies\hspace{2cm}\E{(\rhat_{i}\cdot\rhat_{j})(\rhat_{i'}\cdot\rhat_{j'})}=1/3
\end{align}
The variance therefore reduces to 
\begin{align}
\Var{\langle\RnormSQ\rangle}&=
\sum_{i\ne j}^{J}
\langle \delta_{i}\delta_{j}\rangle
\langle \delta_{i}\delta_{j}\rangle
\frac{1}{3}
+
\sum_{i\ne j}^{J}
\langle \delta_{i}\delta_{j}\rangle
\langle \delta_{j}\delta_{i}\rangle
\frac{1}{3}
=
\frac{2}{3}
\sum_{i\ne j}^{J}
\langle \delta_{i}\delta_{j}\rangle^2
\end{align}
Observing the correlations 
\begin{align}
\langle \delta_{i}\delta_{j}\rangle \equiv \langle \delta_{i}\delta_{i + (j-i)}\rangle\equiv C_{\NoiseVec}(j-i)
\end{align}
depend only on the difference index $k = j-i$, which runs between $-(J-1)$ and $J-1$, we may reexpress the sum over all $i\ne j$ as 
\begin{align}
\Var{\langle\RnormSQ\rangle}&=
\frac{2}{3}
\sum_{k=-(J-1)}^{(J-1)}\left(J-\abs{k}\right)
 (C_{\NoiseVec}(k))^2 - J (C_{\NoiseVec}(0))^2
\end{align}
Since $C_{\NoiseVec}(k) = C_{\NoiseVec}(-k)$, we may reexpress the above simply as a one sided sum
\begin{align}
\Var{\langle\RnormSQ\rangle}&=
\frac{4}{3}
\sum_{k=1}^{J-1}\left(J-k\right)
 (C_x(k))^2
\end{align}
where we have removed the contribution $J(C_{\NoiseVec}(0))^2$ by commencing the sum at $k=1$.

\bla
\newpage
\section{Generating Aribtrary PSDs from Fourier Synthesis}\label{Appendix:FourierSynthNoise}
\bla

\noindent
Following previous work on engineering noise processes~\cite{SoarePRA2014}, we construct the $x(t)$ as a superposition of phase-randomized cosines
\begin{align}
x(t) =\frac{\alpha\omega_0}{2}\sum_{q=1}^{Q}qF(q)\Big[e^{i(\omega_q t+\psi_q)}+e^{-i(\omega_q t+\psi_q)}\Big],
\hspace{1cm}
\psi_q\sim\text{UniformDistribution}[0,2\pi]
\end{align}
Here $\alpha$ is a global scaling factor for setting the total power content, $\omega_0$ is the mode separation incrementing the (angular) frequency in the Fourier superposition, and the mode frequencies are given by $\omega_q \equiv \omega_0 q$. The function $F(q)$ specifies the relative weighting of the Fourier components. The autocorrelation function is then given by 
\begin{equation}\label{Eq:FourierACfunction}
C_x(\tau) = \langle x(t+\tau)x(t)\rangle_t=\frac{\alpha^2\omega_0^2}{2}\sum_{q=1}^Q(qF(q))^2\cos(\omega_q \tau)
\end{equation}
\noindent  where $\langle\cdot\rangle_t$ denotes averaging over all times $t$ from which the relative lag of duration $\tau$ is defined. Invoking the Wiener-Khintchine theorem and moving to the Fourier domain we then obtain the power spectral density
\begin{align}
S(\omega) = \frac{\pi\alpha^2\omega_0^2}{2}\sum_{q=1}^Q(qF(q))^2\Big[\delta(\omega-\omega_q)+\delta(\omega+\omega_q)\Big].
\end{align}
Thus in this formulation the power-spectral density is represented as a Dirac comb of discrete frequency components with the amplitude of the $j$th tooth determined by the quantity $(q(F(q))^2$.  It is then straightforeward to specify the construction of any power-law PSD by writing the amplitude of the $q$th frequency component as a power-law, $S (\omega)\propto (q\omega_{0})^{p}$. It therefore follows that the envelope function for the comb teeth in the phase modulation scales as
\begin{eqnarray}
F(q) = q^{\frac{p}{2}-1}
\end{eqnarray}
Table~\ref{NoisePowerLaws} shows the functional form required for $F(q)$ in order to achieve dephasing-noise PSDs of interest.  

\begin{table}[hbp]\label{}
  \centering
\begin{tabular}{c|cccc|}
\cline{2-5}
 & \multicolumn{4}{|c|}{Power Laws} \\
\cline{2-5}
 &$1/f^2$ &  $1/f$ &White &Ohmic  \\
\hline
\multicolumn{1}{|c|}{p}& $-2$ &$-1$ & $0$  &  $1$  \\
\multicolumn{1}{|c|}{$F(q)$}& $q^{-2}$ &$q^{-3/2}$ & $q^{-1}$  &  $q^{-1/2}$\\
\hline
\end{tabular}
\caption{Functional form of $F(q)$ for well-known power-law PSDs}\label{NoisePowerLaws}
\end{table}
\noindent
From the standpoint of the Clifford sequence \emph{sampling} the noise process, the PSD reconstruction $\tilde{S}(\omega)$ is obtained as the band-limited Fourier transform of \ref{Eq:FourierACfunction}, restricted to the domain $\tau\in[-J,J]$. Hence 
\begin{align}
\tilde{S}(\omega)&=\FT\Big\{C_x(\tau)
\left[\Theta(\tau+J)-\Theta(\tau- J)\right]
\Big\}\\
&=
\label{Eq:BandLimitedPSD}
\frac{J\alpha^2\omega_0^2}{2\sqrt{2\pi}}
\sum_{q=1}^{Q}
\left(qF(q)\right)^2
\Big[
\sinc(J(\omega-\omega_j))+
\sinc(J(\omega+\omega_j))
\Big]
\end{align}
where 
\begin{align}
\lim_{J\rightarrow\infty}\tilde{S}(\omega) = S(\omega)
\end{align}

\bla
\newpage
\section{Confidence Bounds for $\hat{\mu}^{(J)}$}
\bla
%

\noindent 
In the main text we have showed the noise averaged fidelity $\FidNoiseAv$ is, to first order,  a random variable specified by 
\begin{align}
\FidNoiseAv=\Delta-\nu
\end{align}
where $\nu\sim\Gamma\left(\alpha,\hspace{0.1cm}\beta\right)$ is a gamma-distributed random variable, and the values of the parameters $\Delta$, $\alpha$ and $\beta$ are dependent on character of the noise. In particular, restricting attention to the Markovian or DC regime, we have
\begin{table}[H]
\centering
\begin{tabular}{|l|c|c|}\hline
       & Markovian                                           & DC                        \\ \hline\hline 
            
$\alpha$ & $\frac{3}{2}n$                                                           & $\frac{3}{2}$                    \\\hline 
            
$\beta$  & $\frac{2}{3}J\sigma^{2}n^{-1}$                                   & $\frac{2}{3}J\sigma^{2}$\\\hline 
            
$\Delta$    & $1+\frac{2}{3}J^2\sigma^4$  & $1$\\           \hline 
\end{tabular}
\caption{Input parameters for calculating PDF of $\FidNoiseAv$.}
\label{AppendixTable:T1}
\end{table}
\noindent The true mean fidelity formally obtained as an expectation over all possible fidelity outcomes $\mathcal{F}$ defined on the support of the random variables $\CliffVec$ and $\NoiseVec$ is given by 
\begin{align}
 \langle\mathcal{F}\rangle_{\CliffVec,\NoiseVec}
\equiv\E{\FidNoiseAv} = \Delta-\alpha\beta
\end{align}
These equivalent expressions for the total expectation value are hereafter denotes simply by $\mathbb{E}[\mathcal{F}]_J$, as in the main text. In the standard RB procedure this expectation value is estimated by the sample mean 
\begin{align}
\hat{\mu}^{(J)}=\frac{1}{k}\sum_{i=1}^{k}\overline{\boldsymbol{F}}^{(J)}_{i,\langle\cdot\rangle}
\end{align}
The mean gate error $p_{\text{RB}}$ is then approximated by the decay constant from an exponential fit for increasing $J$. It is therefore of interest to quantify the reliability of the estimate $\hat{\mu}^{(J)}$ as a function of the ensemble size. This may be expressed in terms of confidence intervals, treating the measured values $\overline{\boldsymbol{F}}^{(J)}_{i,\langle\cdot\rangle}$ as random variables sampled from the distribution describing $\FidNoiseAv$. The distribution of $\hat{\mu}^{(J)}$ is therefore specified by  
\begin{align}
\hat{\mu}^{(J)} = \Delta-\overline{\nu},
\hspace{1cm}
\overline{\nu} \equiv \frac{1}{k}\sum_{j=i}^k\nu_i,
\hspace{1cm}
\nu_i\sim\Gamma\left(\alpha,\beta\right)
\end{align}
where, since the $\nu_i$ are i.i.d. random variables we have $\overline{\nu}\sim\Gamma\left(k\alpha,\hspace{0.1cm}\beta/k \right)$. The probability that $\hat{\mu}^{(J)}$ falls outside the confidence interval $[\mathbb{E}[\mathcal{F}]_J-L,\mathbb{E}[\mathcal{F}]_J+U]$ is therefore given by 
\begin{align}
\ProbFail\equiv
1-\mathcal{P}\left(\mathbb{E}[\mathcal{F}]_J-L\le \Delta-\overline{\nu}\le \mathbb{E}[\mathcal{F}]_J+U\right),
\hspace{1cm}
\overline{\nu}\sim\Gamma\left(k \alpha,\hspace{0.1cm}\frac{\beta}{k} \right)
\end{align}
where $L,U>0$ specify the lengths of the lower and upper error bars centered on the expectation value $\mathbb{E}[\mathcal{F}]_J$. Substituting $\mathbb{E}[\mathcal{F}]_J=\Delta-\alpha\beta$ and directly integrating the PDF over these confidence bounds we obtain 
\begin{align}\label{AppendixEq:FailureProbabilityRG}
\ProbFail(\alpha,\beta,L,U,k) = 1-Q\left[k\alpha,k\left(\alpha-\frac{U}{\beta}\right),k\left(\alpha+\frac{L}{\beta}\right)\right]
\end{align}
where $Q(a,z_0,z_1)$ is the generalized regularized incomplete gamma function. This is defined in non-singular cases by
\begin{align}
Q(a,z_0,z_1)\equiv\frac{\Gamma(a,z_0,z_1)}{\Gamma(a)}
\end{align}
where $\Gamma(a,z_0,z_1)\equiv \Gamma(a,z_0)-\Gamma(a,z_1)$ is the generalized incomplete gamma function, $\Gamma(a,z)$ is the incomplete gamma function, and $\Gamma(a)$ is the Euler gamma function. This expression can be further condensed by scaling the error-bar lengths by 
\begin{align}
L &= G_L(\Delta-\mathbb{E}[\mathcal{F}]_J) = G_L\alpha\beta\\
U &= G_U(\Delta-\mathbb{E}[\mathcal{F}]_J) = G_U\alpha\beta
\end{align}
where the values $G_L,G_U>0$ specify the lengths of the error bars as fractions of the expected \emph{infidelity}. In this framework, the reliability of the estimate may be quantified by the requirement that 
\begin{align}\label{AppendixEq:ConfidenceBoundStatement}
\ProbFail(\alpha,\hspace{0.05cm}
\beta,\hspace{0.05cm}
G_L\alpha\beta,\hspace{0.05cm}
G_U\alpha\beta,\hspace{0.05cm}
k) <\epsilon
\end{align}
where $\epsilon$ is a small fraction. Substituting values of $\alpha$ and $\beta$ into Eq. \ref{AppendixEq:FailureProbabilityRG} appropriate for Markovian (M) and DC regimes, we find 
\begin{align}
\ProbFail^{(M)}& = 1-Q\left[
\frac{3kn}{2},\hspace{0.05cm}
\frac{3kn}{2}(1-G_U),\hspace{0.05cm}
\frac{3kn}{2}(1+G_L)
\right]\\
\ProbFail^{(DC)}& = 1-Q\left[
\frac{3k}{2},\hspace{0.05cm}
\frac{3k}{2}(1-G_U),\hspace{0.05cm}
\frac{3k}{2}(1+G_L)
\right]
\end{align}
For user-defined $G_L$, $G_U$ and $\epsilon$, one may then solve the inequality in Eq. \ref{AppendixEq:ConfidenceBoundStatement} for minimum $k$. Thus one may bound from below the size of the ensemble $k$ necessary to justify the uncertainties quoted for the mean gate errors $p_{\text{RB}}$ obtained from RB.

\end{widetext}
\end{document}